\documentclass{aa}  

\usepackage{graphicx, subcaption}
\usepackage{placeins}
\usepackage{txfonts}
\usepackage{ulem}
\usepackage[colorlinks=true,     linkcolor=blue, citecolor=blue, filecolor=blue, urlcolor=blue]{hyperref}

\begin{document} 
	
	\title{Interaction of Trappist-1 exoplanets with coronal mass ejections: Joule heating, Poynting fluxes and the role of magnetic fields}
    \titlerunning{Interaction of Trappist-1 exoplanets with coronal mass ejections}
    \authorrunning{Elekes et al.}
	
	\author{Filip Elekes
		\and
		Joachim Saur
		\and
		Alexander Grayver
	}
	
	\institute{Institute for Geophysics and Meteorology, University of Cologne,
		Pohligstr. 3, D-50969 Cologne\\ 
		\email{f.elekes@uni-koeln.de; jsaur@uni-koeln.de}
	}
	
	\date{Received MONTH DAY, YEAR; accepted MONTH DAY, YEAR}
	
	
	\abstract
    {Flares and associated Coronal Mass Ejections (CMEs) are energetic stellar phenomena that drastically shape the space weather around planets. Close-in exoplanets orbiting active cool stars are likely exposed to particularly extreme space weather whose effects on the planets are not understood well enough. The terrestrial Trappist-1 exoplanets are excellent targets to study the impact of CMEs on close-in planets, their atmospheres and ultimately habitability.}
	{We aim to better understand the role of planetary magnetic fields in shielding the planet energetically from external forcing. We expand on recent studies of CME-induced Joule heating of planetary interiors and atmospheres by including a magnetohydrodynamic (MHD) model of the interaction.}
    {We study the interaction of CMEs with Trappist-1b and e using time-dependent MHD simulations. We consider magnetic flux rope and non-magnetized density pulse CMEs. We calculate induction heating in the planetary interior and ionospheric Joule heating for various intrinsic magnetic field strengths and CME energies.}
	{Magnetospheric compression is the main driver of magnetic variability.
    Planetary magnetic fields enhance induction heating in the interior although the effect is weaker with flux rope CMEs. Single event dissipation rates with 1-hour CMEs amount to $20$ TW and $1$ TW for Trappist-1b and e, respectively.
    Taking CME occurrence rates into account, annual average heating rates are $\approx 10$ TW (b) and $1$ TW (e), which are placed near the lower end of previously estimated dissipation rates. Within the range of studied planetary magnetic field strengths $B_p$, magnetospheric inward Poynting fluxes scale with $B_p^3$. Thus, stronger magnetic fields increase the absorption of CME energy. Ionospheric Joule heating rates amount to $10^{3-4}$ TW and decrease for stronger magnetic fields $B_p$. These heating rates exceed the average stellar XUV input by $1-2$ orders of magnitude and might severely impact atmospheric erosion. In a steady state stellar wind ionospheric Joule heating amounts to $\approx10^{2}$ TW.
    }
	{}
	
	
	\maketitle
	%
	\section{Introduction}
	In the Solar system, flares and associated coronal mass ejections (CMEs) are the most energetic solar events that drastically affect the space weather around the planets. Flares and CMEs both originate from the energy release caused by a magnetic reconnection in the stellar atmosphere \citep{Compagnino2017,Chen2011,Shibata2011,Chen2006} and thus, CME mass and velocity are generally correlated with the flare energy. So far there have been no fully conclusive observations of extrasolar CMEs, but it can be assumed that such eruptive events are particularly common around cool stars and are much more extreme than is the case in the Solar system \citep{Moschou2019}. Low mass M-dwarfs are magnetically very active and show strong flaring activity with 
    high XRay luminosities exceeding Solar extreme events \citep{Paudel2018,Moschou2019, Seli2021, Yang2023}. From observations of the Sun we know that high flare XRay energies typically correspond to strong CMEs \citep{Youssef2012,Moschou2019}. M-dwarfs are the most common stars in the universe and many exoplanets have been found in such systems. This raises the question what the space weather is like in such systems and how it affects the planet's space environment, atmosphere, and the interior. Another important factor is the role of planetary magnetic fields in shielding the planets and their atmospheres from space weather and whether intrinsic magnetic fields support or hinder habitability \citep{Airapetian2018,Tilley2019,Airapetian2020}. We address the question about the role of intrinsic magnetic fields in regard to electromagnetic shielding by investigating how they influence the absorption and dissipation of energy in the space environment and interior.

	A promising target of studies aiming at 
    environmental effects on exoplanets is the Trappist-1 system hosting seven terrestrial exoplanets. Trappist-1 is a flaring star \citep{Vida2017, Paudel2018, Glazier2020, Howard2023} with three planets residing in the habitable zone \citep{Gillon2017,Payne2024}.
    
    Trappist-1b has no atmosphere \citep{Greene2023} and there are indications of at most a thin atmosphere around Trappist-1c \citep{Zieba2023,Lincowski2023}. The existence of atmospheres on the other planets has not been conclusively clarified and several modeling studies make different predictions on the existence of atmospheres, atmospheric retention and secondary atmosphere production \citep[e.g.;][]{VanLooveren2024,KrissansenTotton2024,KrissansenTotton2023,Krishnamurthy2021,Dong2018,deWit2018,Bourrier2017}.
    All these models rely heavily on our knowledge about planetary energy budgets and therefore a detailed understanding of possible energy sources to the planetary system is crucial. Especially for cool exoplanet host stars in their pre-main sequence phase stellar wind energy input may provide an important energy supply in addition to high stellar irradiation. Strong interior heat sources 
    may drive geodynamic processes that have been shown to play a key role for the habitability by supporting the carbon cycle due to tectonic subduction \citep{Airapetian2020,Hoening2023}, determining the equilibrium state land-ocean fraction \citep{Hoening2019} or by affecting secondary atmosphere compositions through outgassing and volcanism \citep{Tosi2017,Godolt2019,Airapetian2020,Unterborn2022}. In general, star-planet interactions provide many mechanisms for energy exchange between stars, stellar winds and planets. Tidal interactions, for example, can significantly contribute to the interior energy budget \citep[e.g. for Io;][]{Peale1979,Tyler2015,Davies2024}. Modeling studies suggest that Trappist-1 planets might be subject to a strong tidal heating with Trappist-1b exhibiting a heat production rate similar to Io ($\approx 100$ TW up to $1000$ TW in extreme scenarios). Trappist-1c -- e may experience a tidally induced heat production rate similar to the Earth \citep{Luger2017,Barr2018,Dobos2019} (a few TW). \citet{Bolmont2020} calculated tidal dissipation of Trappist-1e in the order of $1$--$10$ TW for a multi-layered planet and several hundred TW for a homogeneous planet. High-energy radiation from the star significantly influences the temperature on the surface or photo-chemical processes in the atmosphere, if present \citep{Tilley2019,Garcia-Munoz2023} or facilitates atmospheric escape \citep{Roettenbacher2017,Airapetian2017,Bourrier2017}. It was also suggested that the planet's motion through a variable stellar magnetic field may impact the interior of exoplanets by Ohmic dissipation \citep{Kislyakova2020}. Varying stellar winds and magnetic fields may also lead to significant Ohmic dissipation (i.e. Joule heating) within the upper atmospheres \citep{Strugarek2024}, which has also been shown for Trappist-1e \citep{Cohen2024}. 
    
    In this paper we address the question whether space weather, more precisely magnetic variability imposed by planet-intersecting CMEs, dissipate energy within close-in rocky exoplanets and their atmospheres. We study the energy dissipation as a function of the associated flare energy and planetary magnetic field strength. We address how the planets' magnetic fields influence the intake of CME energy into the magnetosphere and whether magnetic fields in general shield the planet energetically from its space environment.

	Interior heating of terrestrial planets due to electromagnetic induction has been investigated in previous studies. The planet's motion through a time variable stellar magnetic field has been considered \citep{Kislyakova2017,Kislyakova2018, Kislyakova2020}. These studies, however, assumed favorably inclined stellar magnetic fields and too fast, outdated stellar rotation rates \citep[e.g. Trappist-1;][]{Kislyakova2017}. These studies suggest a high probability of molten or partially molten exoplanets or local magma oceans which might cause a volcanically driven atmosphere. However, recent observations show that Trappist-1b may have no atmosphere \citep[e.g.;][]{Greene2023}.
    A more recent study used a Trappist-1 flare frequency distribution to estimate the occurrence rate of CMEs intersecting the planets \citep{Grayver2022}. The geomagnetic response of the terrestrial magnetic field environment was scaled to CME events associated with flare energies from the observed flare frequency distribution \citep{Paudel2018}. The study showed that CME-induced Ohmic dissipation in planetary interiors represents a permanent heating mechanism that influences interior heat budgets. \citet{Grayver2022} considered Earth-like magnetized and non-magnetized planets and found that planetary magnetic fields amplify the interior heating significantly.\\
    
    In this work we aim to extend the modeling of CME-induced interior heating of \citet{Grayver2022} by considering the full plasma interaction between interplanetary CMEs and planets, taking into account a parameter range to describe intrinsic magnetic fields, and better understand the role of planetary magnetic fields in such star-planet magnetic interactions. 
    We consider flare frequency distributions for Trappist-1, obtained from dedicated observing campaigns \citep{Paudel2018,Howard2023} to estimate the rate of CME events.
    We study the CME-planet magnetic interaction as a function of CME energy and CME 
    energy partition by assessing the electromagnetic energy transfer towards the planetary surface, surface magnetic variability and the resulting Ohmic dissipation in the planet's interior. We model CMEs dominated by mechanical and magnetic energy to assess differences in energy dissipation and conversion. We also consider an O$_2$ dominated atmosphere for Trappist-1e and assess the energy dissipated within its upper atmosphere in terms of ionospheric Joule heating.
	\section{Numerical simulation} \label{section:numerical_simulation}
	In this section we introduce our physical and numerical models to describe the interaction of Trappist-1b and 1e with a surrounding steady state stellar wind and with time-dependent CMEs. 
 
	\subsection{Method}\label{section:method}
	We carry out single-fluid ideal MHD simulations. To this end, we solve the following MHD equations 
	\begin{eqnarray}
		\frac{\partial \rho}{\partial t} + \nabla \cdot\left[\rho \vec{v}\right] &=& Pm_n - L m_p \label{continuity-equation} \\
		\frac{\partial \rho \vec{v}}{\partial t} + \nabla \cdot \left[ \rho \vec{v} \vec{v}  + p- \vec{B} \vec{B} + \frac{1}{2}  B^{2} \right] &=& -(L m_p + \nu_n \rho)\vec{v} \label{momentum-equation} \\
		\frac{\partial E_t}{\partial t} + \nabla \cdot \left[(E_t + p_t)\vec{v} - \vec{B}(\vec{v}\cdot\vec{B}) \right] &=& -\frac{1}{2} (L m_p + \nu_n \rho)v^{2} \nonumber 
            \\ & & -\frac{3}{2} (L m_p + \nu_n \rho) \frac{p}{\rho} \nonumber 
            \\ & & +\frac{3}{2} (P m_n + \nu_n \rho) \frac{k_B T_n}{m_n} \label{energy-equation}\\
		\frac{\partial \vec{B}}{\partial t}- \nabla \times \left[\vec{v} \times \vec{B}\right] &=&0 \label{induction-equation}
	\end{eqnarray}
	where $\rho$ is the mass density, $\vec{v}$ the velocity and $\rho \vec{v}$ the momentum density. $E_t$ is the total energy density, $E_t = \rho e + \rho v^2 /2+ B^2 / 2\mu_0$, and $e$ the specific internal energy, $p_t$ the total pressure (e.g. magnetic and thermal) and $p$ the thermal pressure. $\vec{B}$ is the magnetic flux density. The terms $P$ and $L$ are production and loss related source terms that are introduced by the presence of a neutral atmosphere. The collision frequency between plasma and neutral particles is $\nu_n$. The atmosphere model as well as production and loss terms are introduced in Sect. \ref{section:TauBoötis_model}. The mass of plasma ions and neutrals are denoted by $m_p$ and $m_n$, respectively. The temperature of the neutral species is denoted by $T_n$.
	The system is closed by an equation of state in the form $p = \rho e (\gamma - 1)$, where $\gamma$ is the ratio of specific heats for the adiabatic case.

    In this work we use spherical and Cartesian coordinate defined as follows. The positive x-axis is parallel to the star-planet line. The z axis is perpendicular to the orbital plane and parallel to the planetary dipole and rotational axis in our model. The y-axis completes the right handed coordinate system and points in the direction of orbital motion. The co-latitude $\theta$ is measured from the positive z-axis, longitudes $\Phi$ are measured from the positive y axis within the xy-plane. The origin is located at the planetary center. We note that the star is not part of the simulation domain. 
	\subsection{Stellar wind model}\label{section:stellar_wind}
	\begin{table*}
		\caption{Physical simulation parameters.}      
		\label{table:parameters}      
		\centering                                   
		\begin{tabular}{c c c c c}          
			\hline\hline                       
			&  Symbol & \textbf{Tr-1 b} & \textbf{Tr-1 e} & Source \\    
			\hline\hline
			Planet radius 	& $R_p$ & $6926.9$ km & $5855.2$ km & \citet{Gillon2017, Turbet2020}  \\      
			Orbital period & $P_{orb}$ & $1.51$ d & $6.09$ d &\citet{Gillon2017}  \\
			Semi-major axis & $a$ & $0.011$ au & $0.028$ au &\citet{Gillon2017}  \\

			\hline\hline 
			\textbf{Stellar wind model} & &\textbf{Tr-1b} &\textbf{Tr-1e}&\\
			Therm. pressure & $p_{sw}$ & $1.83$$\cdot10^{-6}$ Pa& $1.01$$\cdot10^{-7}$ Pa & \citet{Dong2018}\\
			Ion density & $n_{sw}$ & $6.59$$\cdot10^{10}$ m$^{-3}$ & $5.79$$\cdot10^{9}$ m$^{-3}$ & \citet{Dong2018}\\
			Velocity & $\vec{v}_{sw}$ & $(-470,80,-1)$ km s$^{-1}$ & $(-604,50,3)$ km s$^{-1}$ &  \citet{Dong2018}\\
			Magnetic flux density & $B_{sw}$ & $(381,81,-147)$ nT & $(-149,13,-42)$ nT & \citet{Dong2018}\\  
			\hline\hline                    
		\end{tabular}
        \tablefoot{
        Details of the stellar wind and planet model are discussed in Sects. \ref{section:stellar_wind} and \ref{section:TauBoötis_model}. Vectorial quantities are given in Cartesian coordinates where the x-axis is parallel to the the star-planet line. The z-axis is parallel to the orbital plane and planetary magnetic moment. The y-axis completes the right-handed coordinate system.
        }
	\end{table*}
	We apply time-independent stellar wind boundary conditions to the in-flow boundary at the upstream hemisphere ($\Phi = 0$ to $180^\circ$). Stellar wind plasma parameters were taken from \citet{Dong2018} and are summarized in Table \ref{table:parameters}. For all simulations we chose the stellar wind model with maximum total pressure and obtained the plasma parameters from \citet{Dong2018} as the planets are exposed to this wind regime for most of the time according to the model. The stellar wind predictions of \citet{Dong2018} have large uncertainties due to the unknown magnetic field of Trappist-1 and uncertain free parameters of the wind model. Other stellar wind predictions exist that suggest, for example, sub-Alfvénic conditions near most of the Trappist-1 planets and stellar wind velocities on the order of 1000 km/s \citep{Garraffo2017}. The choice of the stellar wind model is expected to have strong effects on the simulated magnetospheres of the Trappist-1 planets. The focus of this work is however on the effects of CMEs, whose energy densities considered in this study significantly exceed those of the steady state stellar wind. The orbital motion of the planets is included in the relative velocity $\vec{v}_0$. The stellar wind plasma consists purely of hydrogen ions.
	\subsection{Trappist-1b and e model}\label{section:TauBoötis_model}
	
	\noindent
	We chose the Trappist-1b and e planets as targets for our study. At Trappist-1b the CME-planet interaction is most energetic due to the proximity to the star and Trappist-1e is of particular interest for atmosphere and habitability studies and is the target of numerous ongoing and future JWST campaigns. In our basic model we assume the planets to not possess any atmospheres. This assumption is most likely true for Trappist-1b, corroborated by newest JWST secondary eclipse observations \citep{Greene2023}. The existence of an atmosphere on Trappist-1e is neither proven nor refuted.

	Therefore we also study the effect of thin atmospheres on magnetic variability and CME energy dissipation within and around Trappist-1e. For this purpose we implement a radially symmetric upper atmosphere composed of molecular oxygen O$_2$. The proposed low density of Trappist-1e may indeed indicate a substantial amount of H$_2$O present within its mantle and crust \citep{Barr2018} serving as a source of atmospheric O$_2$. Hydrogen and oxygen are the primarily lost neutral species due to XUV induced atmospheric loss and photo-dissociation of surface H$_2$O \citep{KrissansenTotton2022,Bourrier2017}. Therefore we include plasma production by photo-ionization of atomic oxygen.
    
    We apply a radially symmetric barometric law for the neutral density. With the radial distance from the planet's center, $r$, and an assumed scale height $H=0.06\,R_p$ (in order to sufficiently resolve the atmosphere in our grid), the neutral particle density can be calculated as
	\begin{equation}\label{eq:atmosphere}
		n_{\text{O}_2}(r) = n_{\text{O}_2,0} \exp\left(\frac{R_p - r}{H}\right) \;,
	\end{equation}
	where the base density of $n_{\text{O}_2,0} = 8\times 10^{6}\,\text{O}_2\, cm^{-3}$ is assumed. With this base density a shell with an increased plasma density is created above the planetary surface and incoming plasma is nearly brought to a halt through the collisions of the plasma with the neutrals and the slow, newly produced ions by photo-ionization. Thus, larger base densities would not enhance the interaction strength, i.e. not further slow the flow near the planet (see \citet{Saur2013} for a discussion on the interaction strength as used here).

	We apply a photo-ionization rate of $\beta_{ph}=6.43\times 10^{-5}\,\text{s}^{-1}$ that results, based on analytical considerations, in the atomic oxygen mass loss rate estimated by \citet{Bourrier2017}, $\dot{M}_{\text{O}} = 5.7\times10^7$ g s$^{-1}$ supplied by the O$_2$ atmosphere. In appendix \ref{Sect:AtmMassLoss} we elaborate on our approach to obtain the photo-ionization rate.
    The photo-ionization production term in Eqs. \ref{continuity-equation}--\ref{energy-equation} is defined by 
    \begin{equation}\label{eq:production_term}
        P(\vec{r},t) = \beta_{ph}\, n_{\text{O}_2}(\vec{r})\; .
    \end{equation}
    The neutral atmosphere interacts with the plasma through ion-neutral collisions with collisional cross section $\sigma_c = 2\times 10^{-19}$ m$^2$ \citep[e.g.;][]{Johnstone2018, Duling2014}. The frequency of collision between ions and neutrals, $\nu_n$, is on the order of $\nu_n\approx 1\,\text{s}^{-1}$ so that
    \begin{equation}\label{eq:collisionFrequency}
        \nu_n = \sigma_c \, n_{\text{O}_2}(r)\, \Bar{v}\;,
    \end{equation}
    where $\Bar{v} \approx v_{sw}$ is a characteristic velocity.
    Photo-ionization is absent in the planet's shadow. In the full shadow (i.e. umbra) ionization is set to a minimum value of $0.1 \cdot \beta_{ph}$ to mimic electron impact ionization on the night side. Since the electron temperature and density on the night side is not known, this minimum value is a guess. A similar ratio is sometimes assumed in solar system research for photo-ionization dominated atmospheres \citep[e.g.;][]{Strack2024}. From the umbra terminator towards the half shadow (i.e. penumbra) terminator $\beta_{ph}$ increases linearly towards the basic value. The sub-stellar point is approximately equal to the upstream side (i.e. near the x-axis). We note that the neutral species is not simulated and not altered by the interaction, it only affects the production, loss and deceleration of plasma. Dissociative recombination of oxygen ions with free electrons is calculated using\\
    \begin{equation}\label{eq:loss_term}
        L(\vec{r},t)=\alpha_r n_{\text{O}_2}(\vec{r})(n_{\text{O}_2}(\vec{r})-n_{sw})\;,
    \end{equation}
    where $\alpha_r$ is the recombination rate. Recombination is switched off when the plasma density falls below the background stellar wind density. The recombination rate of O$^+$ is assumed to be $\alpha_r=5\times10^{-14}$ m$^3\,$s$^{-1}$ for an ionospheric electron temperature of 2500 K \citep{Christensen2012, Walls1974}. We note that in the range 1500 -- 3000 K the variation of the rate coefficient is minor, thus, the value is applicable as a first approximation although the electron temperature near Trappist-1e is not known. 
    
    For simulations without an atmosphere the production and loss terms, P and L, as well as the collision frequency $\nu_n$ in Eqs. \ref{continuity-equation}--\ref{energy-equation} are set to zero.\\

    Although only tentative observations of exoplanetary magnetic fields exist \citep{Turner2021}, three of four terrestrial planets in our Solar system have or had intrinsic magnetic fields. Therefore, our study also includes an investigation of the roles of exoplanetary magnetic fields on the CME-planet interaction and CME energy dissipation within the planets. We assume purely dipolar planetary magnetic fields with magnetic moments parallel to the planetary rotation axis, and perpendicular to the orbital plane. To reduce the computing time, equatorial strengths between $B_p = 0$ G and $B_p = 0.21$ G are used (for reference, Earth's equatorial magnetic field strength is $\approx 0.4-0.5$ G). 
	\\
	The planetary magnetic fields are implemented using the insulating-boundary method by \citet{Duling2014} which ensures that magnetospheric currents do not close within the planetary surface.
	
	\subsection{Interplanetary CME models}\label{Sect:CME_models}
	In this section we introduce our CME models, CME injection into the simulation domain and the simulation procedure.

    In this work we consider two types of CMEs: Density pulse (DP) CMEs solely contain mechanical energy due to velocity and density enhancements. Flux rope (FR) CMEs contain magnetic flux ropes and have enhanced velocity. With this choice we can study in detail the transfer and conversion of CME mechanical and magnetic energy to magnetic variability and which role planetary magnetic fields play during the conversion. Consequently, we do not restrict our study to a single CME type as the nature of extrasolar CMEs is currently unknown. The description of the models can be found in the subsequent sections \ref{Sect:PulseModelCME} and \ref{Sect:FluxropeModelCME}.

	Due to computational cost of time-dependent simulations we constrain the CME event duration to 1 hour. The external and internal timescales, i.e. background plasma convection time and Alfvén time within the magnetosphere, include the magnetosphere's full response to the changes in the upstream plasma conditions.

	We constrain the CME total energy density by assuming an associated flare bolometric energy of $E_{bol} = 10^{31}$ erg for the basic model and by using appropriate scaling laws for CME parameters obtained from solar system based flare-CME association studies \citep{Aarnio2012,Patsourakos2017,Kay2019}. According to the flare frequency distribution of Trappist-1 flares with bolometric energies of $10^{31}$ erg occur roughly once per day \citep{Howard2023}. For our basic model we use this energy to estimate the CME mass using the scaling law of \citet{Aarnio2012}
	\begin{equation}\label{eq:CME_mass}
		M_{CME} = 2.7 \times \left( \frac{E_{bol}}{100} \right)^{0.63} \mathtt{ g}\;,
	\end{equation}
	where $E_{bol}$ (erg) is divided by 100 to give the approximate XRay energy contained in the bolometric energy following \citet{Guenther2020}. The estimated mass is then used to calculate the CME velocity according to the scaling law \citep{Kay2019}
	\begin{equation}\label{eq:CME_velocity}
		v_{CME} = 660 \log{M_{CME}}-9475 \mathtt{\,km/s}\;.
	\end{equation}
    In Table \ref{Table:CMEParameters} we summarize the CME parameters used in this study.
	Density pulse CMEs are solely characterized by enhanced stellar wind density and velocity, whereas FR CMEs include an intrinsic twisted magnetic structure together with a bulk velocity enhancement. Observational evidence from the Solar system suggests that all interplanetary CMEs should have a flux rope structure, but depending on the location of the observer, they may miss the flux rope structure completely \citep{Song2020,Jian2006} since a flux rope does not necessarily fill the entire CME structure. In this case the planet may only experience the non-magnetized part of the CME which is resembled by our DP model.

    The CME parameters described above correspond to a CME shortly after ejection from the stellar corona. We propagate the CMEs to the planetary orbits by using CME evolution parameters summarized in \citep{Scolini2021}. In general, CME parameters are functions of distance traveled by the CME. The heliocentric distance from which we evolve the CME parameters is $D_0$ which is typically closely above the corona. From $D_0$ the CME plasma parameters follow a scaling law of the form $q(D) = q_0\,(D/D_0)^\alpha$, where $q$ is either velocity $v$, density $\rho$ or magnetic field $B$ and $D$ is the distance from $D_0$. We assume $D_0$ to be at $D_0 = 10\,R_{star}$. From there the velocity decays according to $v(D) \propto D^{0.05}$ and the magnetic field according to $B(D) \propto D^{-1.6}$ \citep{Scolini2021}. 
    For a better comparison, we choose the FR and DP CME to have the same total energy density. Therefore, the DP CME mass density $\rho_\text{DP}$ is calculated by equating the DP kinetic energy density, $(\rho_\text{SW}+\rho_\text{DP})v_\textbf{sw}^2 /2$, with the combined FR CME kinetic and magnetic energy density, $\rho_\text{SW}v_\textbf{sw}^2 /2 + B_\text{FR}^2/2\mu_0$, where the densities of both CMEs are offset by the background stellar wind mass density $\rho_\text{SW}$.
	
	\subsubsection{Density pulse CME model}\label{Sect:PulseModelCME}
	We model the density pulse CME by enhancing the background stellar wind plasma parameters according to a Gaussian profile constrained by the CME front, $x_{front}$, and rear,  $x_{rear}$, position along the x-axis. The length between these two positions is defined by the maximum CME velocity (\ref{eq:CME_velocity}) and the event duration of 1 hour. The DP model profile reads
	\begin{equation}\label{eq:DP_profile}
		q(x) = q_{0}+ q_{max}\cdot\exp \left(-\frac{1}{2}  \frac{\left(x-x_{c}\right)^2}{\hat{D}^2}\right)\;,
	\end{equation}
	where $q$ is either the plasma density $\rho$ or the velocity \text{magnitude} $v$. $q_{0}$ is the steady state stellar wind value and $q_{max}$ is the maximum enhancement of the given quantity. $\hat{D}$ is the characteristic decay parameter defining the shape of the Gaussian curve. Outside the CME (i.e. $x>x_{front}$ and $x<x_{rear}$) the $q(x)$ is set to $q(x) = q_0$, thus outside the CME the steady state stellar wind conditions are assumed as described in Table \ref{table:parameters}.
    The value of $\hat{D}$ is determined by finding a CME profile that fills the space between CME rear and front position. At the front and rear Eq. \ref{eq:DP_profile} amounts to approximately $q_{0}+q_{max} 10^{-3}$. $x_{c}$ is simply the midpoint between the rear and front.
	
	\subsubsection{Flux rope CME model}\label{Sect:FluxropeModelCME}
	We use the non-linear, force-free uniform twist Gold \& Hoyle flux rope CME model \citep{GoldHoyle1960} to describe the interplanetary magnetic flux rope. The magnetic field components in flux rope-centered cylindrical coordinates are
	\begin{eqnarray}\label{eq:Fluxrope}
		B_r &=& 0\\
		B_\phi &=& \frac{T r}{1 + T^2 r^2} B_0 \\
		B_z &=& \frac{1}{1 + T^2 r^2} B_0\;,
	\end{eqnarray}
	where $T$ is the twist parameter, $r$ the distance from the flux rope axis and $B_0$ the maximum magnetic flux density along the axis. The twist, $T = 2 \pi n/l$, depends on the axis length $l$ and on the number of turns $n$ along the axis \citep{Wang2016}. The axial length is measured between the two foot points on the star and ranges from $l=2\, L$ (axis parallel to CME-star line) to $l=\pi \, L$ (flux rope axis connects via a full circle to the star), where $L$ is the heliocentric distance of the CME's leading part from the star. In our simulations we set $L$ to the semi-major axis of the respective planetary orbit (Table \ref{table:parameters}). At 1 AU $l$ is approximately 2.6 AU \citep{Demoulin2016}. We assume $l=2.6 \,L$ in our simulations. Due to the lack of knowledge about $l$ in extrasolar environments we assumed the average between the minimum and maximum $l$. However, the choice of $l$ has a very insignificant effect on the structure of the modeled flux ropes in the planet's vicinity. The number of turns $n=10$ is chosen so that a clear helical magnetic field structure can be seen on the cross-sectional width of the modeled CMEs. This value lies within the range of observed solar FR CMEs \citep[e.g.;][]{Wang2016}.

	We obtain the axial magnetic flux density $B_0$ by estimating the flux rope's magnetic helicity using a scaling law \citep{Patsourakos2017,Tziotziou2012},
	\begin{equation}\label{eq:Helicity}
		\log{H_m} = 53.4 - 0.0524\, (\log{E_{bol}})^{0.653} \exp{\left( \frac{97.45}{\log{E_{bol}}}\right)}\;.
	\end{equation}
	From the estimated magnetic helicity $H_m$ we calculate the axial magnetic flux density $B_0$ using the solution of the magnetic helicity integral given by the following equation by  \citet{Dasso2006},
	\begin{equation}\label{eq:Helicity_GoldHoyle}
		\frac{H_m}{L} = \frac{\pi B_0^2}{2 T^3}\left[\ln\left(1+T^2R^2\right)\right]^2 \;,
	\end{equation}
	where $R$ is the radius of the flux rope. We choose $R$ in such a way that the CME event experienced by the planet has a duration of 1 hour, therefore $R$ is determined by the CME velocity (Eq. \ref{eq:CME_velocity}) and event duration. In Table \ref{Table:CMEParameters} we show the estimated flux rope magnetic field strength for different flare energies.

    We assume a southward-oriented FR axis, resulting in a configuration aligned with the planetary magnetic field. Therefore, reconnection, and thus the transfer of magnetic energy towards the planet in the FR scenario is maximum efficient, resulting in upper limit FR contributions to energy dissipation within the magnetosphere. However, we later show that the mechanical interaction is the main driver of magnetic variability near the planet and therefore we do not expect the FR axis orientation to substantially change the results presented in this work.

    \begin{table}[t]
	\caption{Physical CME parameters.}      
	\label{Table:CMEParameters}      
	\centering                                   
	\begin{tabular}{c | c c | c c  }  
        \Large\textbf{Tr-1e}& \textbf{DP}& &\textbf{FR}& \\
		\hline\hline                       
		$E_{flare}$  & $v_{cme}$  & $\rho_{cme}$  & $B_{cme}$  & $v_{cme}$  \\  
        $\left[\text{erg}\right]$ & [km/s] & [$H^+m^{-3}$] & [G]     &[km/s]\\
		\hline
		&     &  &  &  \\                                
		  $10^{29}$   & 1501.4  & 3.47 $\cdot10^6$ & 1.28 $\cdot10^{-3}$ & 1501.4   \\      
		  $10^{30}$   & 1807.98 & 2.99 $\cdot10^7$ & 4.53 $\cdot10^{-3}$ & 1807.98  \\
		  $10^{31}$   & 2114.56 & 1.88 $\cdot10^8$ & 1.33 $\cdot10^{-2}$ & 2114.56  \\
	    $10^{32}$   & 2421.13 & 9.19 $\cdot10^8$ & 3.37 $\cdot10^{-2}$ & 2421.13  \\
		  $10^{33}$   & 2727.71 & 3.61 $\cdot10^9$ & 7.51 $\cdot10^{-2}$ & 2727.71  \\
		\hline
        \Large\textbf{Tr-1b}&\textbf{DP}&&\textbf{FR}&\\
        \hline\hline                       
		$E_{flare}$   & $v_{cme}$  & $\rho_{cme}$  & $B_{cme}$  & $v_{cme}$  \\   
        $\left[\text{erg}\right]$&[km/s]&[$H^+m^{-3}$]&[G]&[km/s]\\
		\hline
		&     &  &  &  \\                                
		  $10^{31}$   & 2290.23  & 3.29 $\cdot10^{9}$ & 6.02 $\cdot10^{-2}$ & 2290.23  \\
          \hline
	\end{tabular}
 
\end{table}
    
	\subsubsection{Simulation procedure and CME injection}\label{Sect:CME_injection}
	We initiate the interplanetary CMEs through time-dependent boundary conditions exerted on the inflow boundary at the upstream hemisphere ($\Phi = 0$ to $180^\circ$). First we run the simulation with constant stellar wind boundary conditions until a quasi-steady state is reached. Then we initialize the CME by superimposing its plasma parameters onto the background stellar wind and upstream boundary conditions. This superposition is confined to x-positions between $-40\;R_p$ and $-420\;R_p$ at the upstream boundary. The CME then propagates towards the planet as the simulation progresses. The rest of the CME is injected through boundary conditions whose fixed parameters follow the CME front position starting at $-40\;R_p$ with the local CME velocity directed parallel to the x-axis. The CME rear, $x_{rear}$, is given by the CME duration (one hour) and maximum velocity. Boundary conditions at boundary cells with $x < x_{rear}$ (behind the rear of the CME) are set to the steady state stellar wind parameters. When the rear of the CME reaches the planet, i.e. $x_{rear} \ge 0$, all in-flow boundary cells inject the steady state stellar wind (Table \ref{table:parameters}).
    With this setup the CME propagates as a structure with length $|x_{front}-x_{rear}|$ along the x-axis across the simulation domain while its lateral extend along y and z reach beyond the simulation boundary. The CME plasma flow is super-fast magnetosonic, therefore a shock front builds up as the CME propagates. This results in CME durations slightly shorter than 1 hour due to plasma compression. Our models produce CME shocks and profiles similar to previous modeling studies \citep{Hosteaux2019,Chane2006} as well as combined observational and theoretical studies \citep{Desai2020, Scolini2021}. We note that in our simulations the pre-CME magnetosphere is not recovered due to lengthy simulation times needed. We found that the significant magnetic variability occurs during the CME peak and thus neglect the slow magnetic topology changes following the magnetosphere's slow and lengthy recovery phase.

    \subsection{On the uncertainties imposed on our CME model}\label{Sect:Uncertainties}
    The empirical relationships between flare energy and CME properties obtained from solar CME observations show a large spread \citep[e.g.;][]{Yashiro2009}. Furthermore, it is not known how these relationships compare to extrasolar CMEs due to the lack of observations. Consequently, the CME parameters derived using the solar flare-CME relations (Eqs. \ref{eq:CME_mass},\ref{eq:CME_velocity},\ref{eq:Helicity} and \ref{eq:Helicity_GoldHoyle}) are merely rough estimates.

    Numerical studies of CMEs launched from magnetically active stars suggest that strong large-scale magnetic fields are capable of efficiently suppressing or slowing down CMEs if the CME energies do not exceed a certain threshold \citep{AlvaradoGomez2018}.
    CMEs launched from polar active regions at high stellar latitudes, however, are significantly more likely to escape the stellar magnetic field due to the open magnetic field structure \citep{Strickert2024}. Polar CMEs have been observed at the Sun, albeit in lower numbers compared to CMEs from lower latitudes \citep{Lin2022,Gopalswamy2015}. CMEs launched from high latitudes are shown to have a tendency to be deflected towards the magnetic equatorial plane \citep{Kay2019}, increasing the likelihood for such CMEs to intersect planets. Furthermore, statistical studies have shown that flares of M-dwarfs indeed more frequently occur at high latitudes \citep{Ilin2021}.

    Recent observations of coronal dimmings during and after flaring activity suggest that cool stars, albeit possessing strong large scale magnetic fields in the kG range like AB Dor, can successfully launch CMEs although it has not been shown conclusively that the observed dimmings correspond to CME material \citep{Veronig2021}.

    Thus overall, M-dwarfs might have fewer and less energetic CMEs as indicated by their flare energies, but the true distribution remains uncertain.
    Therefore, since CME suppression due to large scale stellar magnetic fields may take energy from escaping CMEs, we note that dissipation rates derived in this work are likely upper limits.
    
    \subsection{Numerical model}\label{Sect:Numerics}
    We utilize the open-source code PLUTO (v. 4.4) in spherical coordinates \citep{Mignone2007} to solve the MHD equations. The Eqs. \ref{continuity-equation} -- \ref{induction-equation} are integrated using a approximate hll-Riemann solver (Harten, Lax, Van Leer) with the \textit{minmod} limiter function. The $\nabla \cdot \vec{B} = 0$ condition was ensured by the extended mixed hyperbolic-parabolic divergence cleaning technique \citep{Dedner2002,Mignone2010}.

	The spherical grid consists of 380 non-equidistant radial, 96 angular non-equidistant and 72 equally spaced angular grid cells in $\phi$ and $\theta$ dimension respectively. 
	The radial grid is divided into three regions. From 1 to 1.2 planetary radii ($R_p$) the grid contains 10 uniform cells. After that from $1.2$ to $12$ $R_p$ the next 150 cells increase in size with a factor of $\approx$ $1.01$ per cell. The last 96 cells from 12 $R_p$ towards the outer boundary at $420$ $R_p$ increase gradually with a factor of $\approx 1.02$. 
	We reduce the angular resolution in $\phi$ dimension gradually from the highest resolution near the planetary equator to the poles with a factor of $1.02$ in order to lower the time step constraint on grid cells near the polar axis.
	We chose a simulation regime that large to avoid any interaction of the altered planetary environment and interplanetary CME magnetic structure with the outer boundary.
    \begin{figure}
        \centering
        \includegraphics[width=1\linewidth]{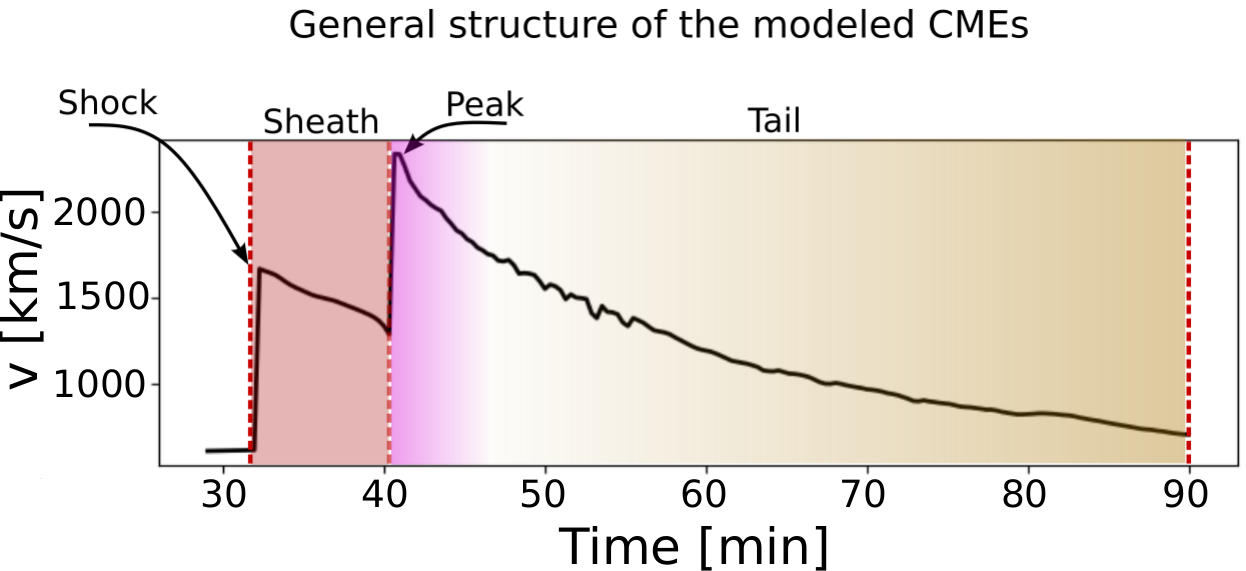}
        \caption{Basic structure of the modeled CMEs. The values are extracted from a fixed location in front of the bow shock.}
        \label{Fig:CME_Structure}
    \end{figure}

    \begin{figure*}

		\centering\includegraphics[scale=0.79]{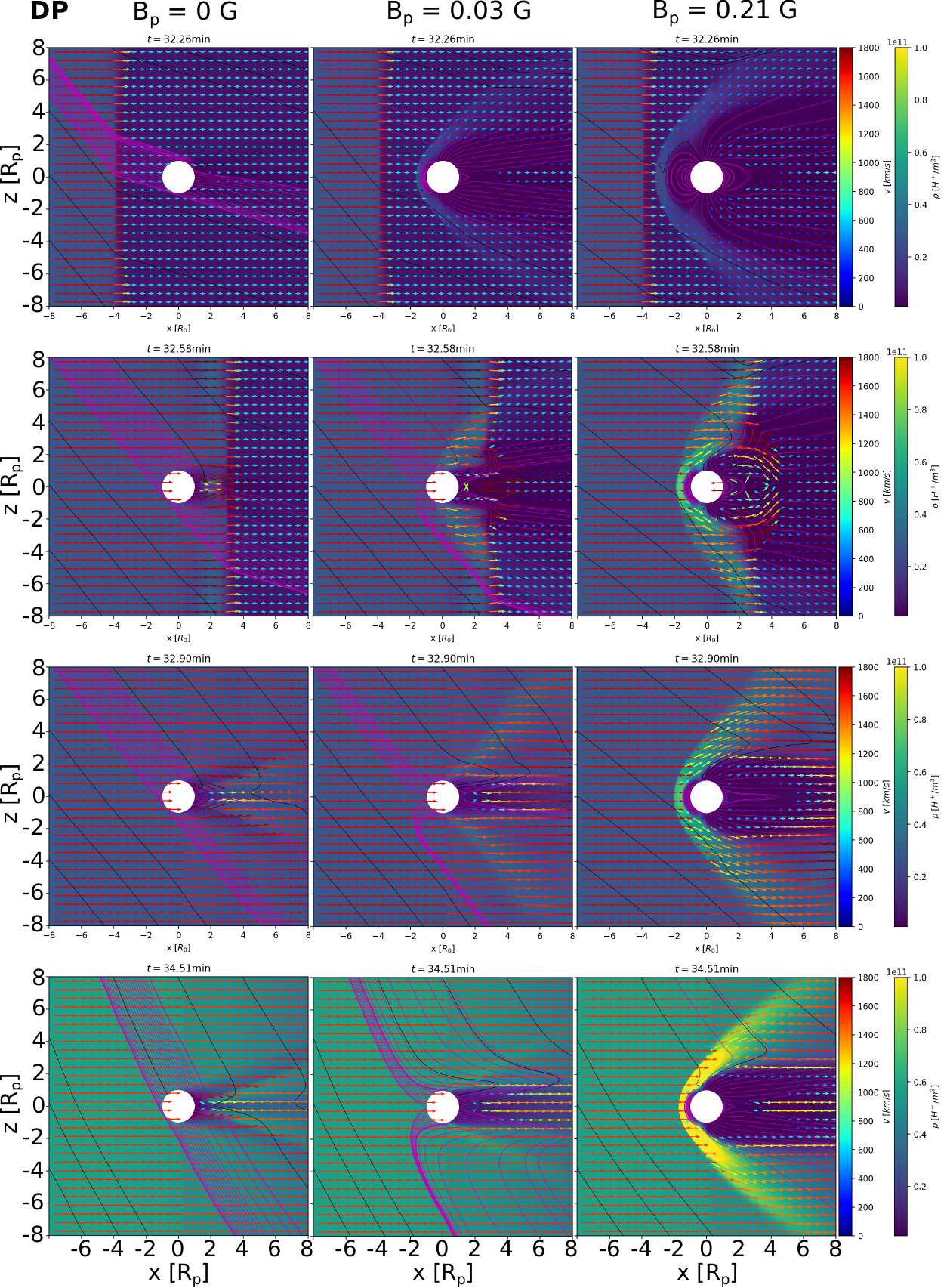}

		\caption{Density pulse (DP) model results. XZ-plane cross sections centered at Trappist-1e (see Sect. \ref{section:numerical_simulation}). Positions are given in planetary radii. Arrows depict velocity vectors and their colors velocity magnitudes (left color bar). Contours indicate plasma density (right color bar). Black and magenta lines indicate interplanetary and planetary magnetic field lines projected onto the xz-plane, respectively. We display the cross sections before the CME shock (top), during the shock crossing (upper middle), during the shath (bottom middle) and during the CME peak (bottom). Each column is for a specified intrinsic magnetic field strength $B_p$. \label{Fig:MSstructure_DP}}
	\end{figure*}

	\begin{figure*}

		\centering\includegraphics[scale=0.79]{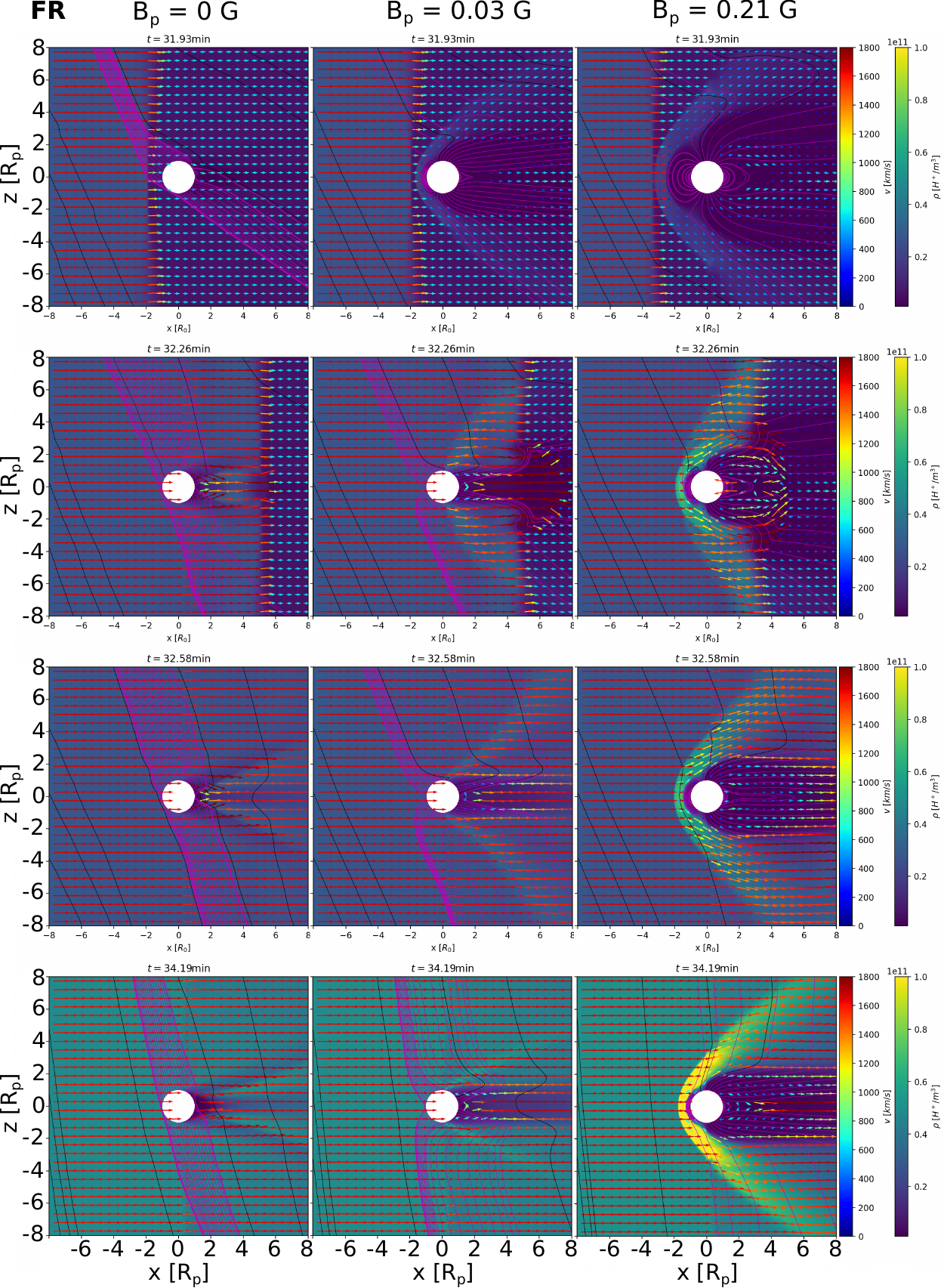}
		
		\caption{Same as \ref{Fig:MSstructure_DP}, but for the  for the Flux Rope (FR) model. Because of a slightly enhanced CME size due to the FR magnetic pressure the CME shock crossing occurs approximately 30 s later compared to the DP case. \label{Fig:MSstructure_FR}}
	\end{figure*}
    
	\section{Results}\label{Sect:Results}
	
	\subsection{Magnetospheric structure}\label{Sect:MagnetosphericStructure}
    \begin{figure}
		\centering
		\includegraphics[width=0.94\linewidth]{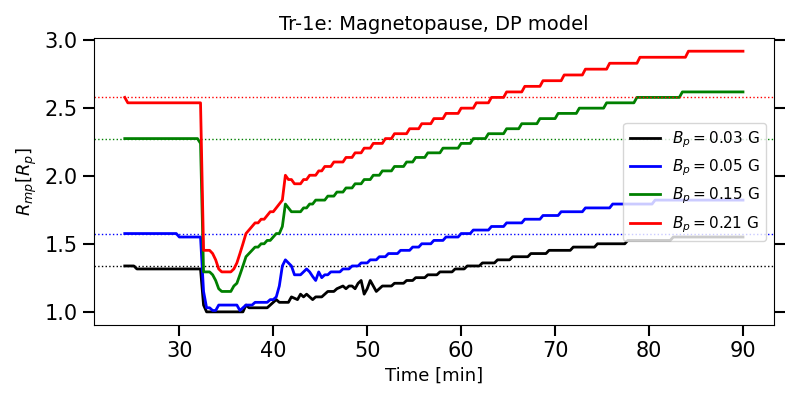}\\
		\includegraphics[width=0.94\linewidth]{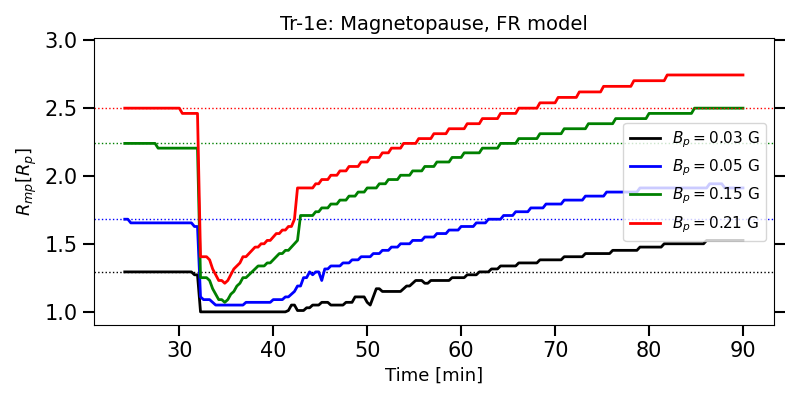}
		\caption{Upstream magnetopause locations (in planetary radii $R_p$) of Trappist-1e for a planetary field with surface strength $B_p=0.03$, $0.05$, $0.15$ and $0.21$ G as a function of time. The upper and bottom panels display DP and FR model runs, respectively. \label{Fig:magnetopause}}
	\end{figure}
	Here we describe the plasma environment and magnetospheric structure during the CME event. In Fig. \ref{Fig:CME_Structure} we show for reference a CME velocity profile at a fixed location in front of the bow shock, letting time evolve, to illustrate the different regions and their names discussed throughout this paper. We additionally show profiles of all plasma quantities in the appendix \ref{Sect:Profiles}, Fig. \ref{Fig:profiles}.\\
    Cross sections through the xz-plane for DP and FR model runs are displayed in Figs. \ref{Fig:MSstructure_DP} and \ref{Fig:MSstructure_FR} respectively. Planetary magnetic fields weaker than $B_p=0.05$ G do not withstand the CME's forcing and the magnetopause is compressed to the planet's surface. From $B_p=0.05$ G to $B_p=0.21$ G the pre-CME magnetosphere has larger sizes from about $R_{mp}=1.4\;R_p$ ($B_p=0.05$) to nearly $R_{mp}=2.6\;R_p$ ($B_p=0.21$). 
    There are no significant differences in the magnetosphere compression between DP and FR CMEs due to the high kinetic energy flux contained in both CME models. A major difference, however, is the magnetic field structure of the FR CME, where an axial component (i.e. z-component) is added to the CME. The stellar wind magnetic field is nearly radial (i.e. parallel to the flow). The CME front compresses the stellar wind field and tends to align it with the shock front, visible in Fig. \ref{Fig:MSstructure_DP}. This gives the otherwise radial stellar wind field a dominant z-component (i.e. roughly parallel to the planet's magnetic moment), which facilitates reconnection between stellar wind and planetary magnetic field. Flux rope CMEs have a purely axial magnetic field in the center with magnetic field components becoming toroidal (i.e. within the x-y plane) towards the FR boundary. In Fig. \ref{Fig:MSstructure_FR} it is visible how the magnetic field's z-component increases as the FR center approaches the planet.
    This enhances the reconnection efficiency between ambient and planetary magnetic field. Released magnetic energy due to reconnection at the downstream magnetopause accelerates plasma towards and away from the planet.
    This can be seen in Figs. \ref{Fig:MSstructure_DP} and \ref{Fig:MSstructure_FR} indicated by high velocities in the planet's plasma shadow at roughly $x>2$ $R_p$ where stream lines diverge. This is best visible for strong magnetic fields like in the $B_p=0.21$ G case as those store more magnetic energy when the field is strongly perturbed (e.g. right panels of Fig. \ref{Fig:MSstructure_FR}).

	We track the upstream magnetopause position during the CME event to examine the mechanical and magnetic forcing on the planetary magnetosphere. First we extract magnetic field profiles along the x-axis for all times. In our simulations, the magnetic field of the CME that piles up within the CME shock is mostly anti-parallel to the z-axis. From the magnetic field profiles we extracted the position of the magnetopause by determining the position of the reversal of the magnetic field and jump in its magnitude. This method yields sufficiently precise results. Magnetopause location time series for DP and FR model runs are displayed in top and bottom panels of Fig. \ref{Fig:magnetopause}, respectively. We note that the small jumps in Fig. \ref{Fig:magnetopause} are due to the grid resolution. For all planetary magnetic field strengths considered the CMEs compress the magnetopause to $R_{mp}=1$--$1.3\,R_p$. For magnetic fields below $B_p=0.05$ G the magnetopause is pushed to the planetary surface (see also Figs. \ref{Fig:MSstructure_DP} and \ref{Fig:MSstructure_FR}). Even in the strong magnetic field case ($B_p=0.21$ G) the magnetopause location drops from approximately 2.5 $R_p$ to about 1.25 $R_p$. The CME sheath region (i.e. the region between the shock and the CME peak, Fig. \ref{Fig:CME_Structure}) exerts the strongest forcing on the magnetosphere due to maximum total pressure. It starts after the shock crossing at 32 minutes and lasts for about 10 minutes. 
    
	In all simulations the magnetosphere undergoes a structural change that remains as long as the CME decays. After the CME main phase a new temporary equilibrium between magnetospheric and CME pressure is reached. 
    During the CME decay phase a tail of diluted low plasma density follows the CME and accordingly decreases the CME kinetic energy while the velocity and magnetic flux density mostly decayed to the steady state values. Consequently the magnetosphere inflates during the CME decay phase.
	
	\subsection{Magnetic variability and interior Joule heating}\label{Sect:InteriorOhmicHeating}
	\begin{figure}
		\centering
		
		\includegraphics[width=0.99\linewidth]{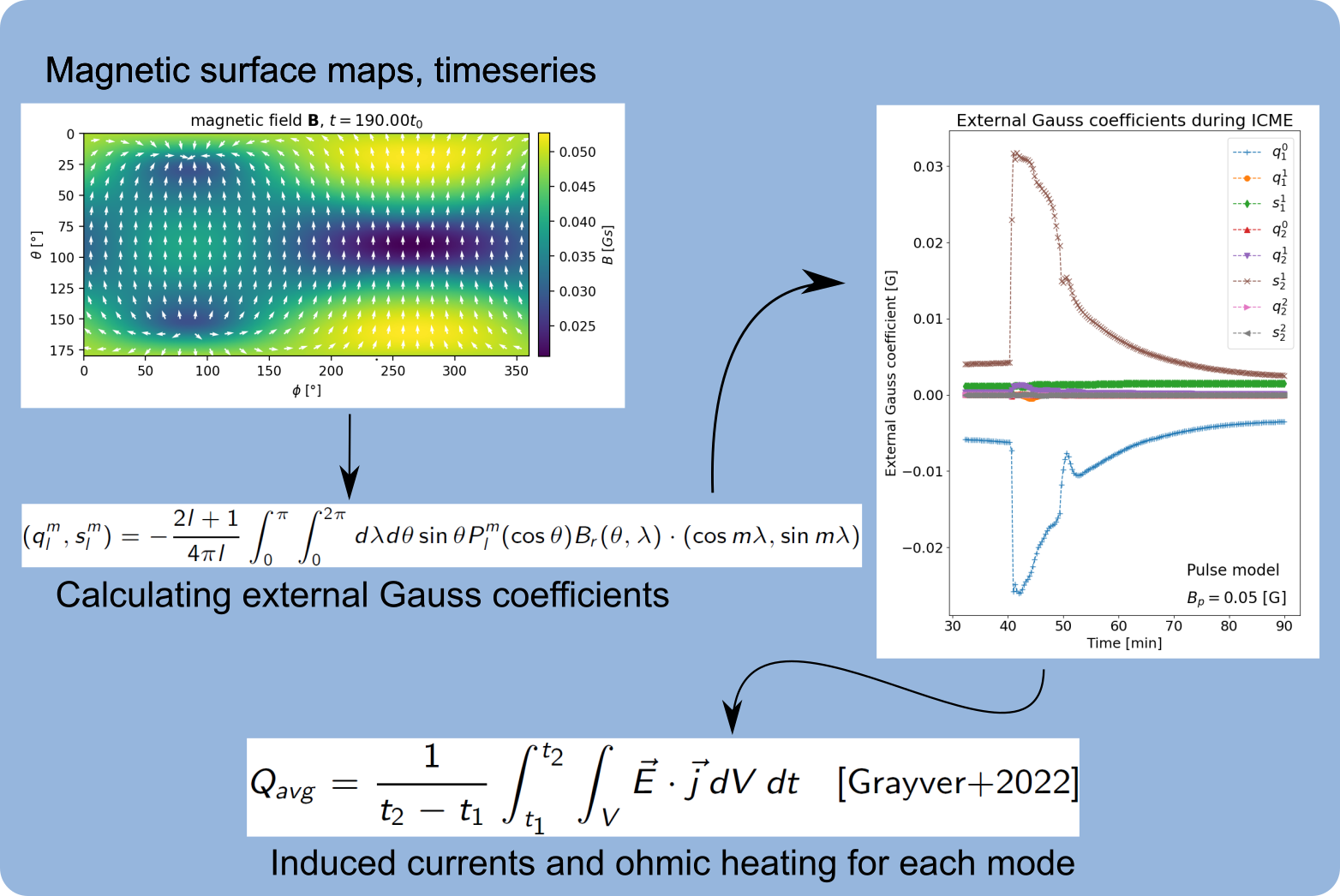}
		\caption{A schematic of the post-processing pipeline to calculate CME induced Joule heating in the interior of a planet. \label{Fig:OhmicHeatingPipeline}}
	\end{figure}

	\begin{figure*}
		\centering
        \includegraphics[width=0.99\linewidth]{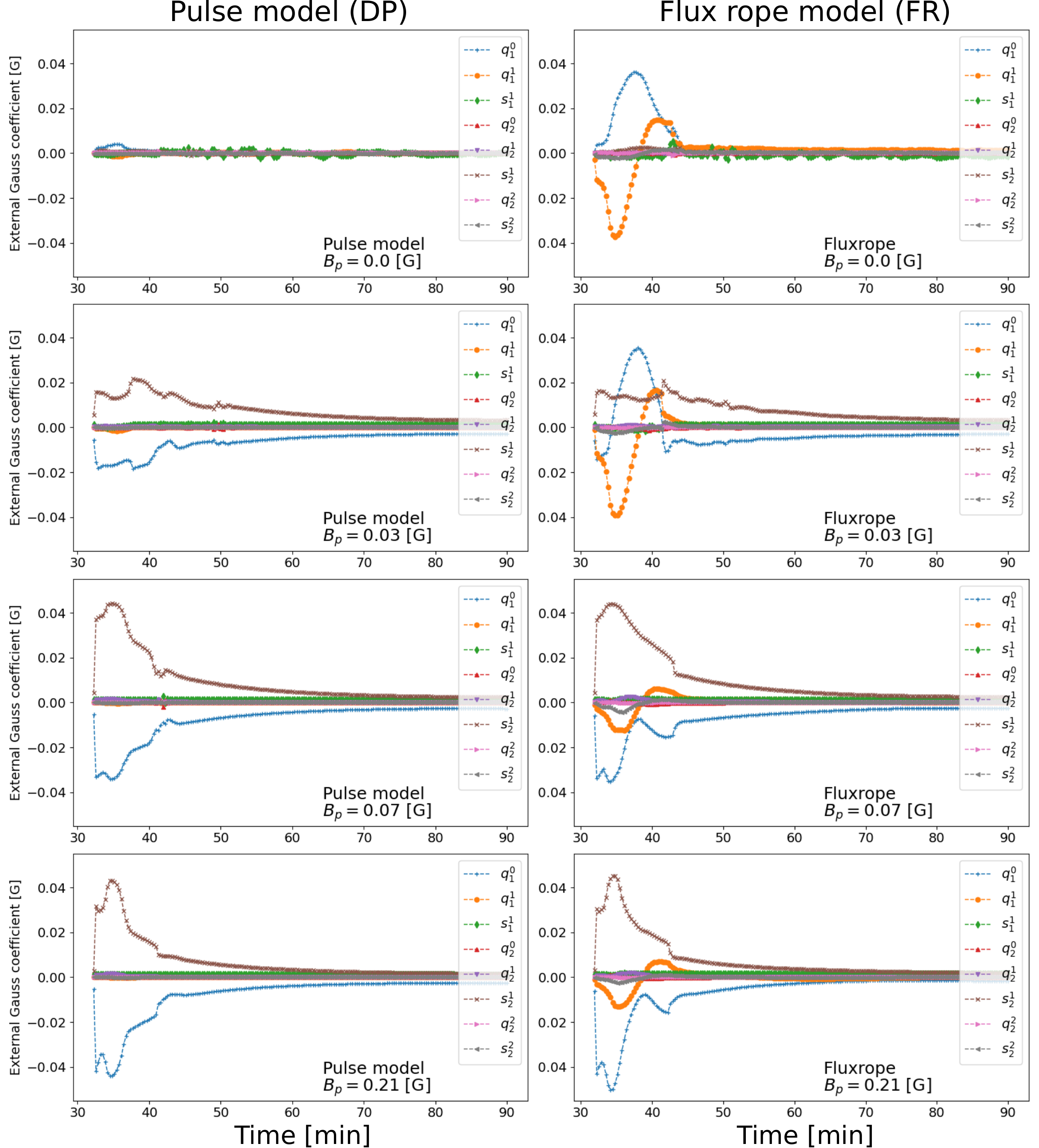}
		\caption{Evolution of the external Gauss coefficient (Eq. \ref{eq:GaussCoeff}) during one CME as function of time (minutes) for planetary magnetic field strength $B_p = 0$ (top), $0.03$ (upper middle), $0.07$ (lower middle) and $0.21$ G (bottom). The left column shows DP, right column FR model results. Displayed are all considered external Gauss coefficients up to quadrupolar degree ($l=2$).\label{Fig:GaussCoefficients}}
	\end{figure*}

    Alongside the structural change of the magnetosphere due to the CME impact, temporal magnetic field variability near the planet is generated by the plasma interaction. Varying magnetic fields $\dot{B}$ induce eddy electric fields in the planetary subsurface by virtue of electromagnetic induction. The resulting electric fields drive electric currents $\vec{j}$ within electrically conductive layers. These induced currents within the planetary subsurface give rise to energy dissipation through Joule heating when electric conductivity is finite. We follow the modeling approach of \citep{Grayver2022} and define the total Joule heating rate within the planetary body as 
    \begin{equation}\label{eq:ohmicHeating}
		Q_{J} = \frac{1}{T}\int_{t_0}^{t_1}\int_{V}^{} \vec{E}\cdot \vec{j}\; dV\,dt\;,
	\end{equation}
    where $T$ is the duration of the CME event, $V$ the planetary volume, $t_0$ and $t_1$ the start and end times of the integration and $\vec{E}$ the electric field.

	To solve Eq. \ref{eq:ohmicHeating}, we first need to describe the magnetic field variability at the planetary surface. In order to do so we extract vectorial magnetic field maps directly above the planet's surface during the CME event and decompose the external field (e.g. the magnetic field without the constant dynamo magnetic field) for each step up to the quadrupolar degree ($l = 2$) using spherical harmonics multipole expansion. We also considered higher degree spherical harmonics which, however, have negligible amplitudes and thus also negligibly contribute contribute to interior heating. This decomposition is valid for potential fields within the upper non-conducting subsurface but we only have access to magnetic field components above the surface. We, however, validated that the magnetic field directly above the surface can be approximated well as potential field by directly comparing the magnetic field at the inner boundary and the unperturbed magnetic field transmitted into the domain by the boundary conditions. The coefficients of the multipole expansion of the field give us information about which magnetic field mode (e.g. dipolar or quadrupolar) that is generated by the interaction is dominating the magnetic field variations at the surface of the planet. We calculate the external Gauss coefficients, 
	\begin{equation}\label{eq:GaussCoeff}
		\left(\begin{array}{c} 
			q_l^m \\ s_l^m
		\end{array}\right) = n(l) \int_{0}^{\pi} \int_{0}^{2\pi} d\lambda d\theta \sin \theta P_l^m (\cos \theta) B_r(\theta, \lambda)
		\left(\begin{array}{c} 
			\cos m \lambda \\ \sin m \lambda
		\end{array}\right)\;,
	\end{equation}
	where $n(l) = -\frac{2l+ 1}{4\pi l}$ is the Schmidt semi-normalization factor, $P_l^m$ are the associated Legendre polynomials of degree $l$ and order $m$. 
    The radial magnetic field components $B_r$ as function of co-latitude $\theta$ and longitude $\lambda$ are extracted from the simulations. We note that this process is done in post-processing and thus, induction in the interior does not couple back to the space environment. The calculations and steps done to obtain interior Joule heating rates are summarized in Fig. \ref{Fig:OhmicHeatingPipeline}.

	External Gauss coefficient time series are shown in Fig. \ref{Fig:GaussCoefficients} for Trappist-1e with magnetic fields of $B_p = 0$, $0.03$, $0.07$ and $0.21$ G and CME-associated flare energy of $10^{31}$ erg. For simplicity we omit Trappist-1b in this discussion due to the similarity of the magnetic fields' dynamical behavior (i.e. the physical behavior is the same but differs only in magnitude).

	Flux ropes have spatially inhomogeneous magnetic fields (see e.g. Fig. \ref{Fig:profiles}), and when convected on the planet, it causes time-variable fields near the surface. Density pulse CMEs on the contrary have approximately constant magnetic fields but carry more mechanical energy than flux rope CMEs due to plasma density enhancement. This incident mechanical energy, i.e., kinetic energy, compresses the planetary upstream magnetic field, which is the main source of magnetic flux variability observed in the DP model results.

	In DP CME simulations, magnetic variability is dominated by the $q_1^0$ (vertical dipole mode) and $s_2^1$ (equatorial quadrupolar mode) coefficients. The decrease of $q_1^0$ corresponds to an increase in north-south magnetic flux density due to magnetosphere compression. The increase of $s_2^1$ is associated with equatorial magnetic field components with minima at the flanks, sub- and anti-sub stellar point. As we increase the planetary magnetic field strength, these coefficients enhance accordingly until reaching the maximum near 0.04 G for $B_p \ge 0.07$ G. This behavior is indicative of increasingly more efficient induction in the magnetosphere due to the stronger field. In the $B_p=0$ G case there is only a slight increase of the $q_1^0$ coefficient that is due to stellar magnetic field draped around the planet. There are small fluctuations in the $s_1^1$ component most pronounced in the $B_p=0$ G case relating to flow instabilities at the flanks of the planet. Variability in $q_1^0$ and $s_1^1$ scales with $B_p$ due to the enhanced inductive response of the plasma in the planet's space environment (Eq.\ref{induction-equation}).

	In FR CME simulations with a non-magnetized planet the dominant Gauss coefficients are $q_1^0$ and $q_1^1$. The increase and decay of $q_1^0$ around 40 minutes directly correlates with the magnetic flux density profile along the FR with its maximum north-south component at the FR axis (Fig. \ref{Fig:profiles}). Similarly the change of the twisted FR horizontal field components (within the xy-plane) is reflected in the sinusoidal variation of equatorial $q_1^1$ coefficient. The equatorial field component of the FR in the xy-plane ($q_1^1$) translated to the planet decreases in strength as the compression of the magnetosphere gains in effectiveness ($q_1^0$, $s_2^1$). Therefore, with strong planetary magnetic fields the FR magnetic field is more efficiently shielded off by the planet's field and both types of CMEs nearly produce the same Gauss coefficients, dominated by magnetosphere compression. 

	The time series of the external Gauss coefficients are used as input for the interior induction heating model of \citet{Grayver2022} where electromagnetic induction and resulting Ohmic heating is calculated for each Gauss coefficient separately. For the whole CME duration we calculate the Joule heating rate within the whole planetary volume V, $Q_{J}$, for all modes according to Eq. \ref{eq:ohmicHeating} \citep{Grayver2022}. The integration is performed from start $t_0$ to end $t_1$ of the CME event and the result is divided by the CME duration $T=1\text{ h}$. With Eq. \ref{eq:ohmicHeating} we then obtain the heating rate during an one-hour CME event. We refer the reader to \citet{Grayver2022} for a detailed description of the induction heating model. We use a simple homogeneous interior model with the constant electrical conductivity of $\sigma = 0.01$ S/m which corresponds to the typical conductivity in the Earth's asthenosphere and lower lithosphere \citep{Naif2021}. This choice is motivated by the proposed interior composition of the Trappist-1 planets similar to that of Earth \citep{Agol2021}. The likely enhanced electrical conductivity at larger depths does not have a strong effect on our assumption of constant conductivity $\sigma = 0.01$ S/m because currents attenuate with depth due to the skin effect. Most heating occurs in the uppermost layers of the planets in depths up to a few 100 km where electrical conductivity may be approximated to first order as spatially constant. 
	\\
    \begin{figure}
		\centering
		\includegraphics[width=0.5\textwidth]{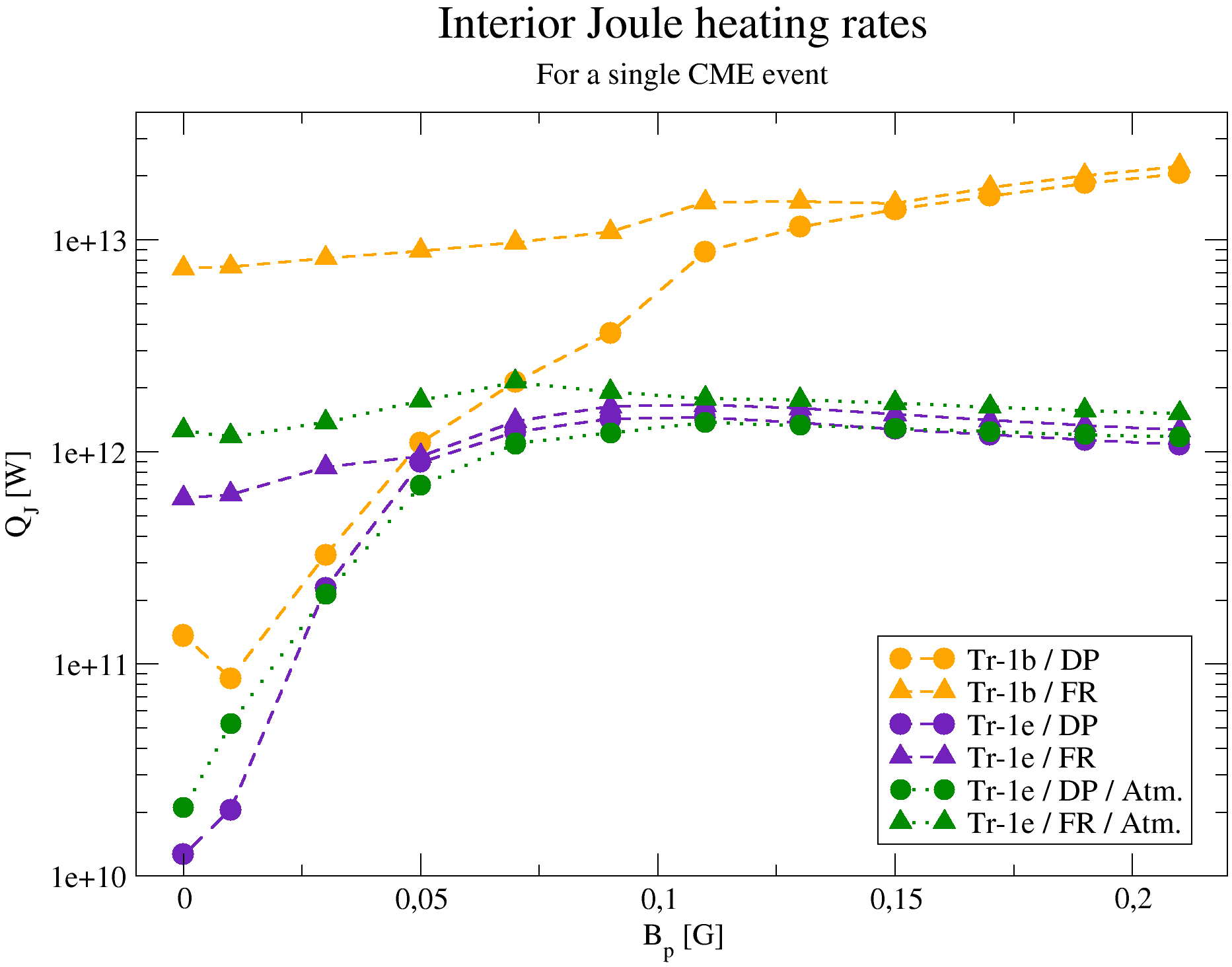}
		\caption{Joule heating averaged within 1-hour CME events in the interior of the planets, $Q_J$ (Eq. \ref{eq:ohmicHeating}), given in Watts as a function of planetary magnetic field strength $B_p$. Triangles denote FR, circles DP CME cases. Yellow data points correspond to Trappist-1e, purple to Trappist-1e. Green data points denote the results with an atmosphere around Trappist-1e.}\label{Fig:HeatingRates}
	\end{figure}

    \begin{figure}
		\centering
		\includegraphics[width=0.49\textwidth]{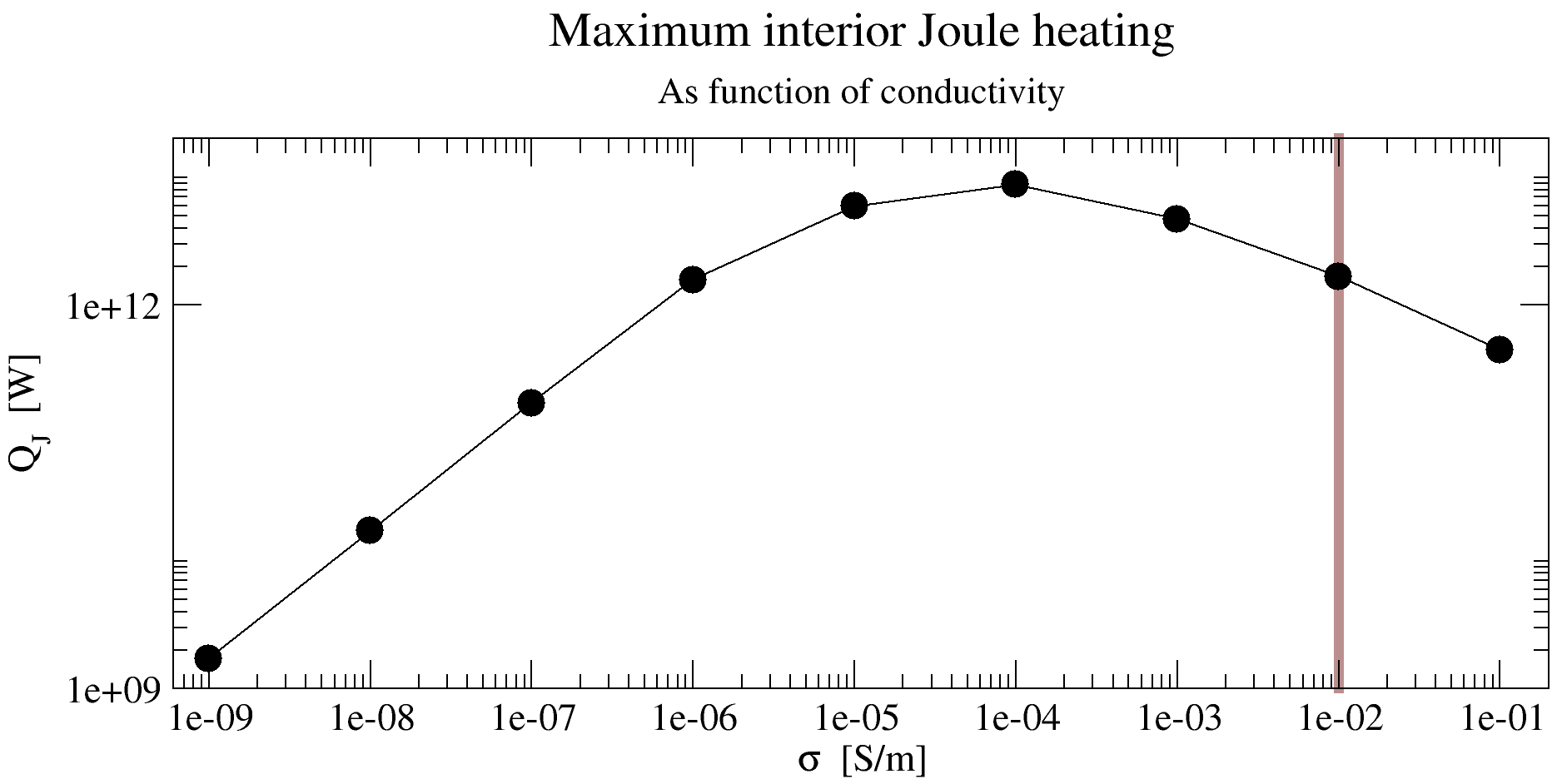}
		\caption{Heating rates at $B_p=0.11$ G for Trappist-1e as a function of the electric conductivity $\sigma$ in S/m. The brown vertical line indicates the homogeneous model conductivity adopted in this study. \label{Fig:HeatingRates_Sigma}}
	\end{figure}

    The dissipation rates $Q_J$ in the interior of Trappist 1b and e for one CME with $E_{flare}=10^{31}$ erg are shown in Fig. \ref{Fig:HeatingRates}. In the $B_p = 0$ G scenario we calculate a heating rate averaged within one hour of approximately $0.01$ TW (Tr-1e) and $0.1$ TW (Tr-1b) in the DP case. CMEs expand during their propagation through the heliosphere and therefore the energy density decreases accordingly. Because of that Trappist-1b experiences stronger CMEs, resulting in higher $Q_J$ as magnetic perturbations enhance with incident energy. For both CME models heating rates reach up to $1$--$2$ TW at $B_p=0.09$ G for Trappist-1e. The maximum heating lies outside our parameter space for Trappist-1b but $Q_J$ flattens towards the maximum $B_P = 0.21$ G of this study towards around 20 TW.
        
    Our calculations show that for the flux rope CMEs the dissipation $Q_J$ is only very mildly dependent on the strength of the internal field $B_p$ (see Fig. \ref{Fig:HeatingRates}). It typically increases as long as $B_p$ is small until $0.1$ G and then maximizes or goes into saturation. This is very different for the density pulse CMEs. The electromagnetic dissipation strongly increases with the strength of the planetary field up to $0.1$ G. The reason is that a stronger $B_p$ enables a stronger generator mechanism to convert the mechanical energy of the CME into electromagnetic energy dissipated within the planet.
    Above  $B_p=0.05$ G (Tr-1e) and $\approx 0.15$ G (Tr-1b) heating rates in both CME models have the same functional dependence on $B_p$ as magnetic variations intrinsic to FR CMEs are increasingly shielded off by the planetary magnetic field (see also Fig. \ref{Fig:GaussCoefficients}).
    
    Interior heating rates for Trappist-1e with the atmosphere considered here are similar to those without such an atmosphere. However, for weak magnetic fields $B_p < 0.1$ G heating rates are slightly enhanced in the FR CME case. An atmosphere weakens the plasma flow near the planet. Therefore mechanically generated perturbations are damped accordingly that results in slightly lower inductive response in the planetary interior. The increase in heating rates in the FR case ($B_p < 0.1$ G) comes from enhanced tension on the reconnected field lines since the motional response of magnetospheric plasma to external forcing is counteracted by the atmosphere that acts as energy sink.\\
    The majority of dissipated energy presented here is dissipated only within the uppermost layers of the planets. This is due to the exponential attenuation of magnetic fields with depth caused by the skin effect.
	\\
	
	To further study the effect of the interior electric conductivity on Ohmic dissipation, we additionally calculated heating rates with conductivities between $10^{-9}$ to $\sigma=0.1$ S/m (Fig. \ref{Fig:HeatingRates}). As the maximum heating rate in the Tr-1e model is achieved at $B_p=0.11$ G we only show the maximum heating rates as a function of the electrical conductivity in Fig. \ref{Fig:HeatingRates_Sigma}. In Fig. \ref{Fig:HeatingRates_Sigma} we see that dissipation maximizes at intermediate conductivities around $10^{-4}$ S/m.
    Therefore, a lithosphere conductivity comparable to that of Earth already produces maximum heating rates in our model. More insulating as well as more conducting lithospheres lead to a lower Ohmic dissipation.
    
    Lastly we note that we also examined the effect of CME duration on heating rates. For durations $T>>1$ h heating rates approach a lower limit approximately a factor of $<2$ smaller than those presented here for $T=1$ h. The effect of CME duration on interior heating is thus, according to our model, insignificant. See appendix \ref{Sect:DurationInfluence} for a discussion on this conclusion.
    
    \subsection{Ionospheric Joule heating}
    In addition to interior Joule heating, energy is also dissipated in the ionosphere through ionospheric Joule heating. The presence of a neutral species introduces collisions between plasma and neutral particles due to their relative motion. These collisions cause the ions to dissipate energy in form of ionospheric Joule heating \citep{Vasyliunas2005}. Ionospheric Joule heating can be calculated using the convective electric field $\vec{E}=-\vec{v}\times\vec{B}$ and the plasma conductivity perpendicular to the magnetic field, i.e. the Pedersen conductivity $\sigma_P$. We calculate the ion term of the Pedersen conductivity for each grid cell which is a function of the ion-neutral collision frequency $\nu_c$, ion gyro frequency $\omega_g$, ion number density $n_i$ \citep{Baumjohann2012},
    \begin{equation}\label{eq:PedersenConductivity}
        \sigma_P = \frac{n_i e^2}{m_i} \frac{\nu_c}{\nu_c^2+\omega_{g}^2}\;,
    \end{equation}
    where $e$ is the elementary charge, $m_e$ the electron mass and $m_i$ the ion mass. The collision frequency is defined as in Eq. \ref{eq:collisionFrequency} and the ion gyro frequency can be calculated with $\omega_g=e B/ m_i$. Ionospheric Joule heating per unit volume can then be calculated with $q_{J,ion}=\sigma_P E^2$. We integrate this expression over the magnetosphere volume $V$ to get the total ionospheric Joule heating rate and integrate the resulting heating rates over the CME event duration. By dividing the result by the CME duration $T$ we get the average ionospheric Joule heating rate for one CME event,
    \begin{equation}\label{eq:IonosphericJouleHeating}
        Q_{J,ion} = \frac{1}{T}\int_{t_0}^{t_1}\int_{V}^{} \sigma_P(\vec{r},t) E(\vec{r},t)^2\;dV\,dt\;,
    \end{equation}
    \begin{figure}
		\centering
		\includegraphics[width=0.49\textwidth]{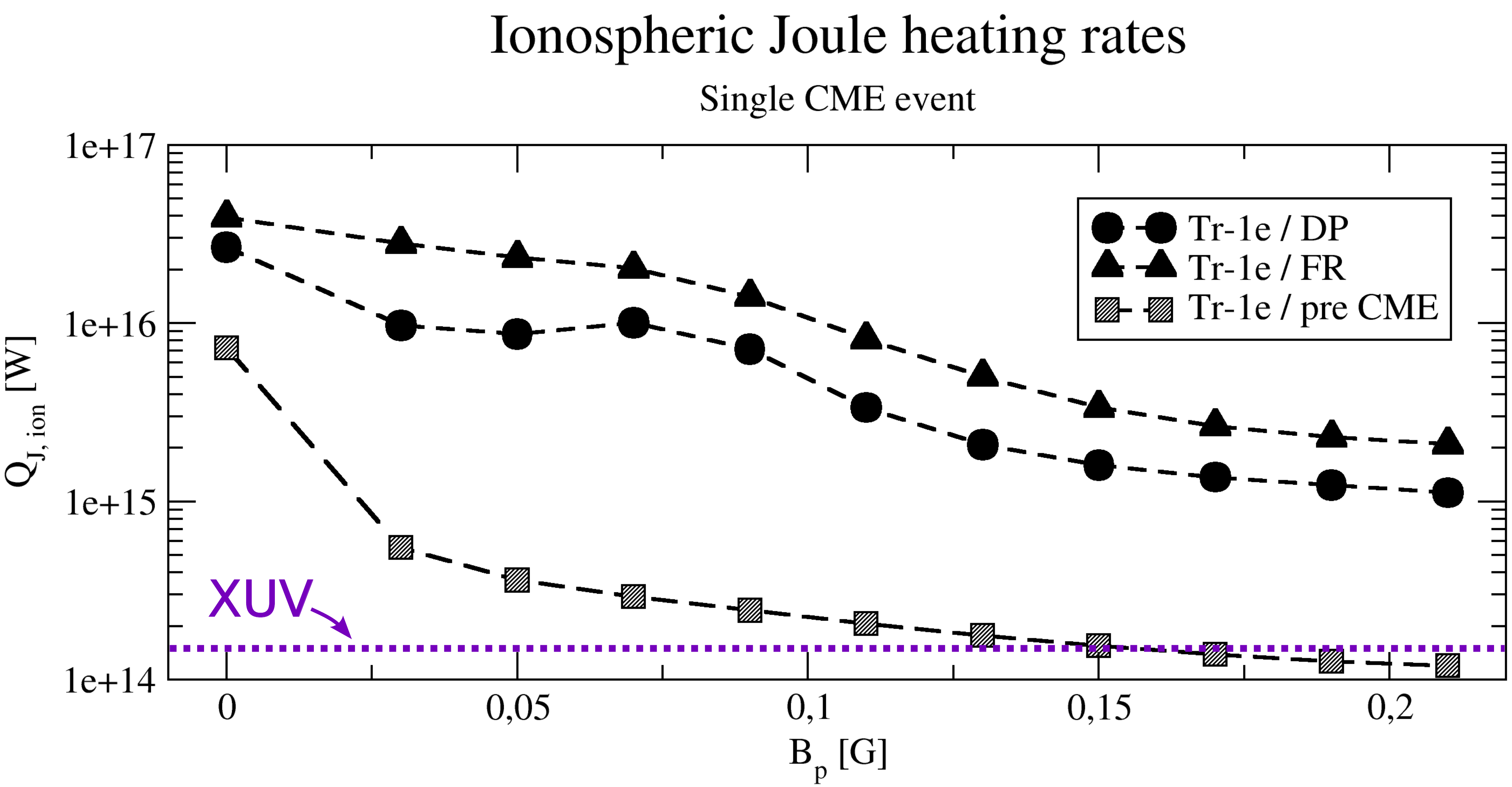}
		\caption{Ionospheric Joule heating rates (Eq. \ref{eq:IonosphericJouleHeating}) averaged within $1$ hour CME events for Trappist-1e as a function of planetary magnetic field $B_p$. Density pulse (DP) and flux rope (FR) model results are denoted by circles and triangles, respectively. The dayside XUV power received from the star is also indicated. \label{Fig:IonosphericJouleHeatingRates}}
	\end{figure}
    where $t_0$ and $t_1$ are the start and end time of the CME event. We integrate Eq. \ref{eq:IonosphericJouleHeating} from the surface at $1\,R_p$ to $1.5\,R_p$ to avoid including the shock and unshocked CME plasma. We only consider magnetic fields $\ge 0.05$ G due to the magnetopause being pushed to the planetary surface for weaker fields. Resulting heating rates are shown in Fig. \ref{Fig:IonosphericJouleHeatingRates}. For DP model runs we obtain heating rates from $3\times10^{4}$ TW ($B_p=0.0$ G) down to $1-2\times10^{3}$ TW ($B_p=0.21$ G) while heating rates in the FR case are approximately a factor of $2$ higher. Contrary to interior Joule heating, ionospheric Joule heating rates decrease with increasing magnetic field strength, because the ionospheric Pedersen conductivity is inversely proportional to the magnetic field strength. 
    \begin{figure*}
		\centering
        \includegraphics[width=0.958\linewidth]{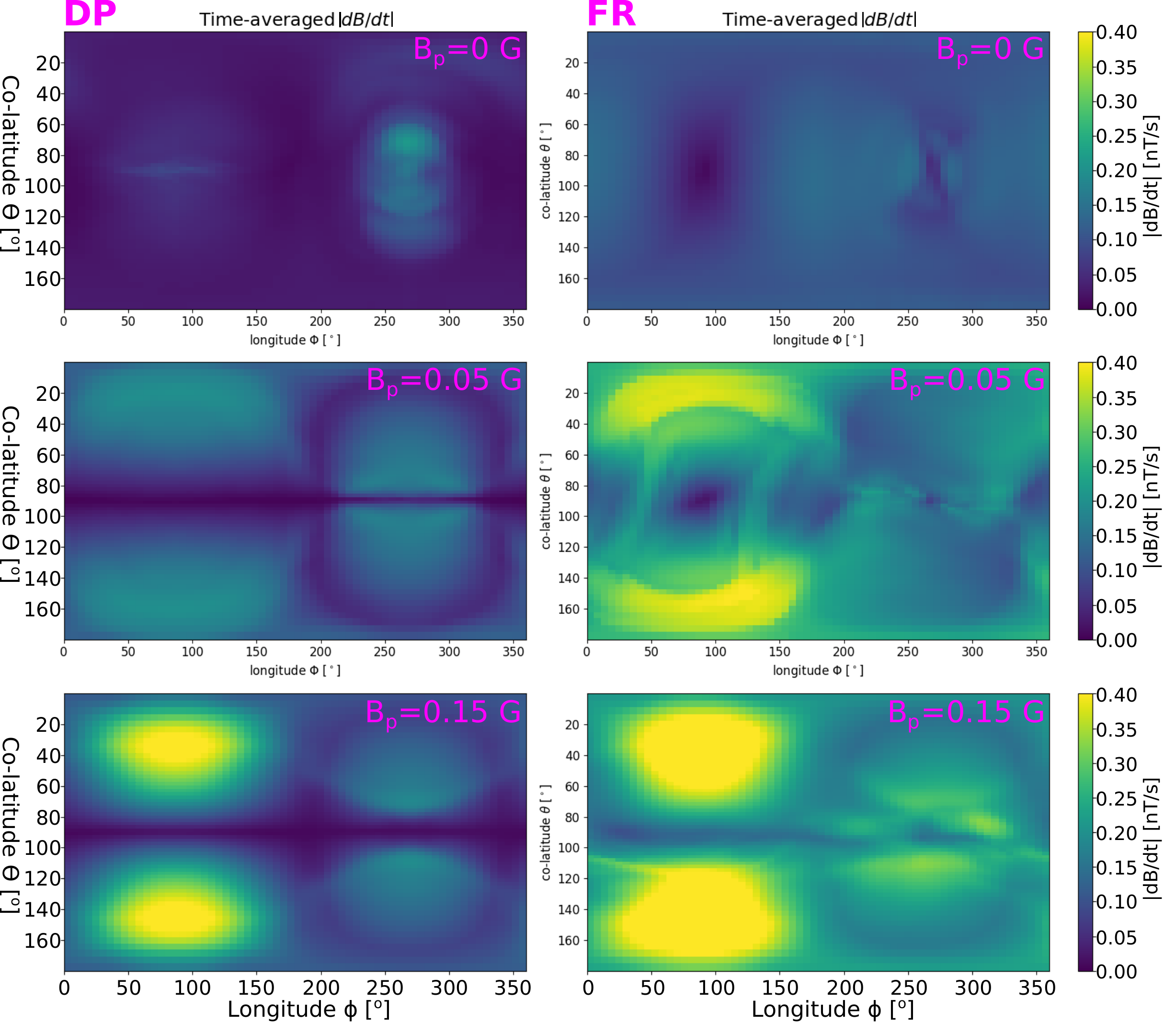}
		\caption{Maps of time-averaged magnetic variability, $dB/dt$ (nT/s), of the radial magnetic field component obtained directly above the planetary surface. Co-latitude and longitude are shown on the Y and X axes, respectively. Upstream direction is at longitude below 180 degrees. The left column shows density pulse (DP), the right column flux rope (FR) model results. We show maps for $B_p=0$ (top), $0.05$ (middle) and $0.15$ G (bottom). \label{Fig:MagneticMaps}}
	\end{figure*}
    In contrast to the interior Joule heating, the effect of Joule heating on ionospheric plasma is nearly instantaneous and about 4 orders of magnitude stronger. Therefore, CME induced ionospheric Joule heating directly heats up the space plasma which may lead to comparably severe effects on the upper atmosphere like atmospheric inflation and significant escape rates during CME events.
    In Fig. \ref{Fig:IonosphericJouleHeatingRates} we additionally show ionospheric Joule heating rates due to the steady state stellar wind before the CME event, $Q_{0,ion}$. In the non-magnetized case the dissipation rates reach up to $8\times10^{3}$ TW and then quickly decay towards $10$ TW as we increase the planetary magnetic field strength. Thus, a CME enhances ionospheric dissipation rates by 1--2 orders of magnitude on average.

    To put these dissipation rates into context we estimate the amount of stellar XUV power received by the dayside of Trappist-1e. XUV flux is the main driver of photo-evaporation and thus has a severe impact on the survivability of planetary atmospheres. Modeling studies based on observations from \citet{Wheatley2017} suggest that the XUV luminosity $L_{XUV}$ of Trappist-1 has saturated at a value of about $L_{XUV}/L_{bol}\approx 10^{-3}$ during a large part of the star's lifetime \citep{Birky2021,Fleming2020}. We take this value as the upper limit XUV luminosity and the bolometric luminosity of Trappist-1, $L_{bol}=5.53\times10^{4}\,\text{L}_\sun$ \citep{Ducrot2020}. We then calculate a total XUV power received by dayside of Trappist-1e of $Q_{XUV}=9.5\times10^1$ TW. Additionally we use the XUV flux from the MEGA-MUSCLES spectrum \citep{Wilson2021} to also estimate the XUV power at the dayside of Trappist-1e similar to \citet{Strugarek2024}. The resulting XUV power is $Q_{XUV}=1.64\times10^2$ TW. We display the upper limit XUV power estimate in Fig. \ref{Fig:IonosphericJouleHeatingRates} for comparison with the ionospheric Joule heating rates.
    
    According to our estimates and model results the planet approximately receives the same amount of XUV power and steady state ionospheric Joule heating power, $Q_{XUV}\approx Q_{0,ion}$, for $B_p > 0.1$ G. During a CME event, however, dissipation rates in the ionosphere quickly increase by 1--2 orders of magnitude, depending on the planetary magnetic field. The fast response of ionospheric plasma to heating rates on the order of 100-1000 TW possibly results in severe atmospheric escape during CME events which makes close-in planets around active stars less likely to retain their atmospheres as they receive this energy in addition to flare XUV fluxes and stellar quiet XUV radiation. We lastly note that our calculated dissipation rates are time-averaged over the duration of the CME and that peak heating during the CME main phase exceeds the presented values considerably.

    Lastly we note that the ionosphere may shield the planetary surface from external magnetic variations due to the skin effect \citep[e.g.;][]{Strugarek2024}. For this shielding to be effective, however, large ionospheric conductances and a high frequency of magnetic variability are needed to decrease the skin depth to lower values compared to the extend of the ionosphere. By height-integrating the Pedersen conductivity (Eq. \ref{eq:PedersenConductivity}) in our simulated ionospheres we find the Pedersen conductance to be approximately 5-20 S in more strongly magnetized cases (from about $0.05$ G) and up to 600 S in the weakest magnetic field case. With the frequency of magnetic variability on the order of one minute we find the skin depth for moderately to strong magnetized planets to be on the order of hundred kilometers. This poses lower limits since the period of magnetic variability in our simulations is considerably longer (see Fig. \ref{Fig:GaussCoefficients}). In the weakest magnetic field cases the skin depth decreases towards tens of kilometers. Therefore, in the case of efficient magnetic screening, the energy dissipated in the ionosphere would increase significantly. However, most importantly our study shows in case of a conductive ionosphere, that the overall dissipation rate in the ionosphere can be orders of magnitude larger compared to the dissipation due to induced fields in the interior. Thus additional shielding in the ionosphere only further increases the dominance of the ionospheric heating.

 	\subsection{Localization constraints on interior heating}\label{Sect:Localization}
	Induction driven interior Joule heating is proportional to the temporal change of magnetic flux, i.e. $Q \propto dB/dt$ whereas only the radial component of the magnetic field is required to assess $Q$ (see Eq. \ref{eq:GaussCoeff}). The plasma interaction between planet and space environment is fundamentally asymmetric since the orientation of the magnetic field and the ambient plasma velocity field are key variables that deform and align the magnetosphere with the flow and ambient magnetic field. In Fig. \ref{Fig:MagneticMaps} we show maps of time-averaged radial component of $|dB/dt|$ over the surface of Trappist-1e with several planetary magnetic field strengths. The trends we describe in the following are also observed in Trappist-1b simulations. We therefore omit these results for simplicity.

    If the planet is non-magnetized most $dB/dt$ is found downstream near the equator due to the high tension on the stellar wind and CME magnetic field lines within the wake of the planet. In the FR scenario $dB/dt$ is generally more homogeneously distributed. As we switch on the planetary magnetic field we observe the average $dB/dt$ to increasingly focus at the upstream hemisphere near the polar cusps due to the planetary magnetic field lines being mostly radial there. For planetary magnetic fields $B_p \le 0.07$ G the average $dB/dt$ in the FR scenario are more widely distributed and washed out over the upstream hemisphere. For stronger planetary magnetic fields the distributions of $dB/dt$ in both CME models become increasingly similar.
    The average $dB/dt$ peak in both CME scenarios at planetary magnetic fields of $B_p\approx0.11$-$0.15$ G. Afterwards the $dB/dt$ decay slightly but remain focused at the upstream polar regions.

	These results suggest that the degree of local focusing of the heating correlates strongly with the planetary magnetic field strength. Stronger magnetic fields favor CME-induced interior heating to peak on the upstream hemisphere at high latitudes.
	\begin{figure}
		\centering
		\includegraphics[width=0.35\textwidth]{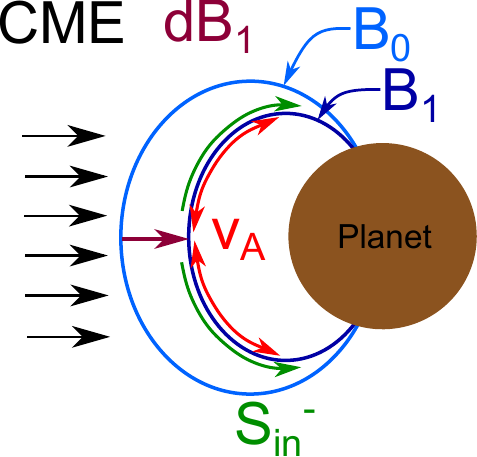}
		\caption{This sketch illustrates the generation of background magnetic field aligned Poynting fluxes associated with Alfvén waves. We do not show fast mode waves due to the negligible effect on the radial components of surface $dB/dt$ found with our model.} \label{Fig:PoyntingAflvenSchematic}
	\end{figure}
	
	\section{Discussion}
	In this section we aim to better understand the electromagnetic energy fluxes around the planets and how they relate to CME-induced interior heating. We first discuss the energetics of the CME-magnetosphere interaction in Sect. \ref{Sect:Energetics}. We  discuss the time variability of near-surface Poynting fluxes during the CME event and assess the energy transfer from CME to the magnetosphere (Sect. \ref{Sect:Energetics_timedomain}). In Sect. \ref{Sect:Energetics_Scaling} we consider time averaged Poynting fluxes, compare them to interior Joule heating and study the scaling of input power near the surface as function of planetary magnetic field strength. In Sect. \ref{Sect:Heating_vs_E} we briefly discuss the dependence of interior heating on CME-associated flare energy. 
    
	\subsection{Energetics of the CME-magnetosphere interaction}\label{Sect:Energetics}
	Magnetic variability is mostly controlled by electromagnetic energy being propagated by Poynting fluxes. In our model these Poynting fluxes are generated by the CME-planet plasma interaction and ultimately deliver the electromagnetic energy to the planet's surface where it is, to some extent, dissipated by Joule heating. We distinguish between two regimes of electromagnetic energy transfer. 
    The closed field line region (i.e. the magnetosphere) is defined by all planetary magnetic field lines that intersect the planet's surface twice. Here magnetic energy is propagated mostly along field lines. The open field line region is defined by highly mobile magnetic field lines that originate in the planet and connect to the stellar magnetic field. Stellar wind / CME energy and mass are injected into the magnetosphere across the open magnetic field. Magnetic energy transport in this regime is dominated by convection.

    Figure \ref{Fig:PoyntingAflvenSchematic} shows a schematic illustrating magnetic variability and associated Poynting fluxes. Before the CME the planetary magnetic field $\vec{B}_0$ is in equilibrium with its surroundings. This perturbed dipole is compressed on the upstream side and elongated on the downstream side (e.g. top panels of Figs. \ref{Fig:MSstructure_DP} and \ref{Fig:MSstructure_FR}). A CME or any variation in the interplanetary medium perturbs the steady state field towards a new, temporary equilibrium magnetic fields $\vec{B}_1$. The perturbation magnetic field $\delta \vec{B}$ (i.e. the residual field $\vec{B}_1 - \vec{B}_0$) drives Alfvén waves perpendicular to $\delta \vec{B}$ which propagate the magnetic variability, $dB/dt$, parallel to $\vec{B}_1$ with Alfvén velocity $v_A$ towards the planet as well as away from it. These Alfvén waves carry electromagnetic energy that is associated with Poynting fluxes $\vec{S}_{||}$ parallel to $\vec{B}_1$. The parallel Poynting flux can be calculated using the residual magnetic and associated electric field,
	\begin{equation}\label{eq:S_parallel}
		\vec{S}_{||} = \frac{\delta \vec{E}\times \delta \vec{B}}{\mu_0} = \frac{\delta B^2}{\mu_0} \vec{v_{A}}\;,
	\end{equation}
	where the velocity $\vec{v}_A=\vec{B}/\sqrt{\mu_0 \rho}$ is the Alfvén velocity \citep{Saur2021,Saur2018,Park2017,Saur2013,Keiling2009}.
    The maximum $dB/dt$ can be expected where $\vec{B}_1$ is perpendicular to the surface of the planet, $|\vec{B}_1| \approx B_r$ (i.e. the field lines are purely radial), so that $S_{||} \approx S_r$.
	\\

\begin{figure}
		\centering
		\includegraphics[width=0.499\textwidth]{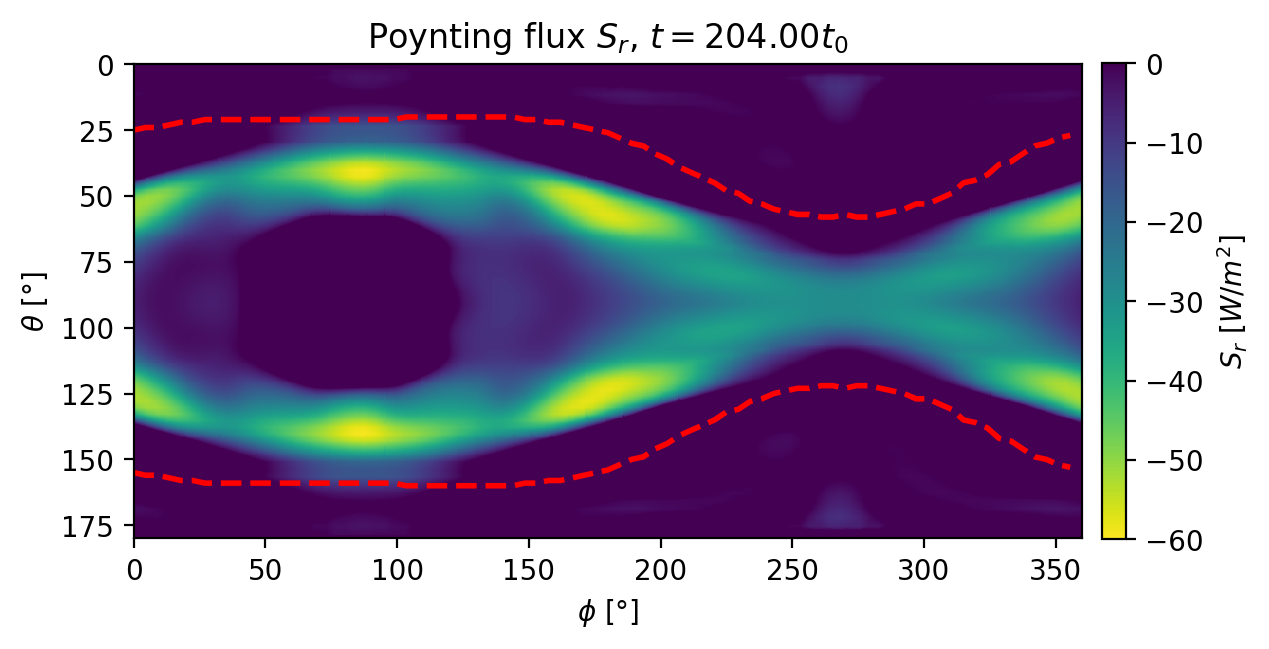}
		\caption{The open-closed field line boundary (red lines) plotted on top an inward Poynting flux, $S_{in}$, map for Trappist-1e, $B_p=0.21$ G, DP model. As an example, the map is shown during the CME sheath crossing, but the boundary remains almost constant over the entire simulation period, for DP as well as FR simulations.} \label{Fig:ocfb}
	\end{figure}

	In Fig. \ref{Fig:ocfb} we show a map of $S_{r}$ directly above the surface of Tr-1e. Only Poynting fluxes towards the planet are shown which results in the negative signs. The red line denotes the open-closed field line boundary (OCFB). For latitudes below the OCFB the field lines are closed. The OCFB shown here is similar in all magnetic field as well as CME models and it changes only minimally during the CME event. 
    Figure \ref{Fig:ocfb} shows that almost all Poynting fluxes $S_{r}$ lie within the closed field line region indicating that the electromagnetic energy parallel to the magnetic field is generated almost exclusively within the closed magnetosphere due to magnetosphere compression. 
    The field lines near the OCFB are most mobile in a sense that they are convected downstream by the external plasma flow. Given the magnetic variability maps in Fig. \ref{Fig:MagneticMaps} we expect the magnetic variability to be mostly transported along magnetic field lines and the distribution of $S_{r}$ within the OCFB supports this. We also show $S_{r}$ maps for $B_p=0.05$, $B_p=0.15$ and $B_p=0.21$ models during the CME main and decay phase in Fig. \ref{Fig:S_map} to illustrate the general dominance of $S_{r}$ in regions where most magnetic variability (i.e. $dB/dt$, Fig. \ref{Fig:MagneticMaps}) is found independent of the CME model. Furthermore the Poynting flux distribution near the CME peak (middle column) correlates best with the magnetic variability maps, indicating that most heating might be generated during the CME sheath and not during the shock crossing (see also Fig. \ref{Fig:CME_Structure}).\\
    We note that magnetic perturbations are also carried by fast mode waves dominantly perpendicular to the magnetic field. However, as the radial magnetic field component is crucial for the induction in the interior we omit the discussion of fast mode perturbations.
	\\
	
	Due to these considerations we focus our analysis in the following sections on the electromagnetic energy that is delivered inward by radial Poynting flux components, $S_{r}$.
    We formulate the integrated Poynting flux as integral over radial Poynting fluxes directed towards the planet,
	\begin{equation} \label{eq:PoyntingFluxInward}
		S_{in}^- = \int_{A_{planet}} S_r^- dA\;,
	\end{equation}
	where $A_{planet}$ is the surface area of the planet and $S_r^-$ are the radial components of the Poynting flux directed towards the planet. Due to the spherical coordinate system the inward flux is negative, therefore we omit the sign and indicate the inward components with the minus superscript.
    In the following sections we use $S_{in}^-$ to study temporal evolution of interaction generated magnetic variability as function of CME model and time and how planetary magnetic fields affect the transfer of magnetic energy towards the planet surface.

	\subsubsection{Magnetospheric Poynting fluxes during the CME}\label{Sect:Energetics_timedomain}
	\begin{figure*}
		\includegraphics[scale=0.45]{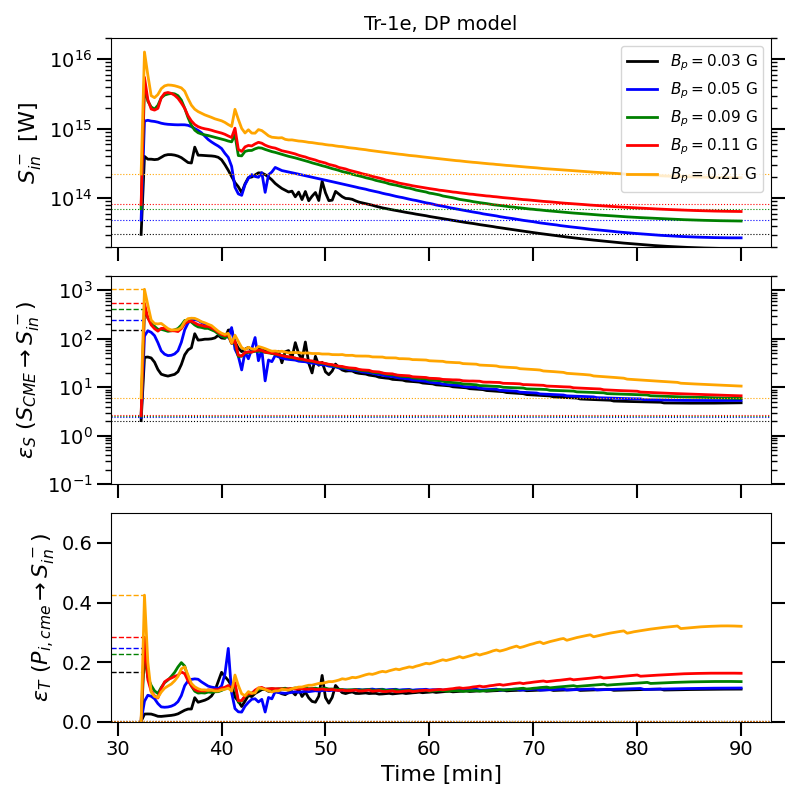}
		\includegraphics[scale=0.45]{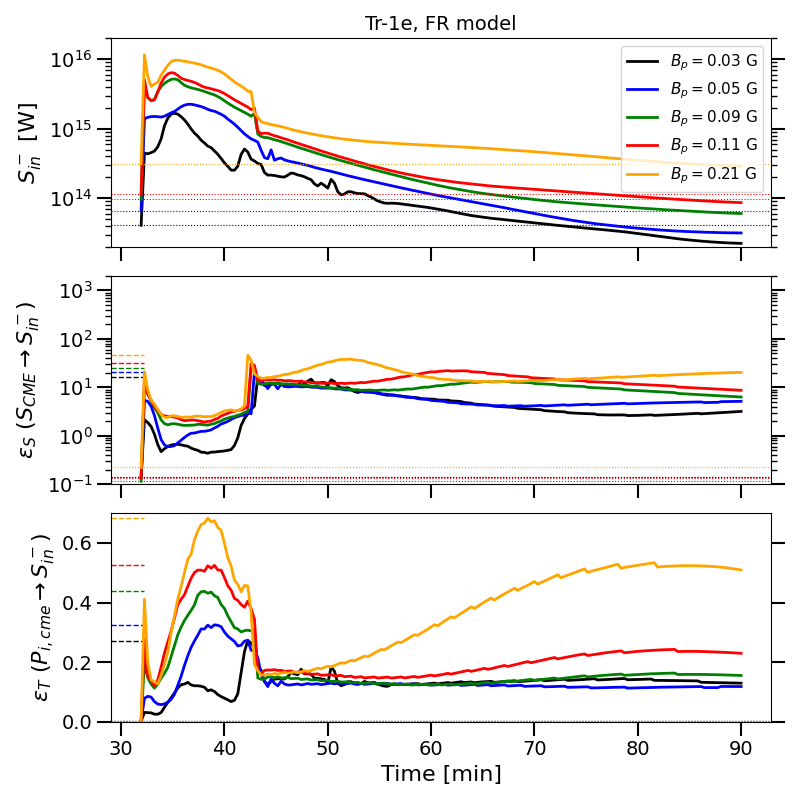}\\
		\caption{Time series of $S_{in}^-$ in W (top), $\epsilon_S$ (middle) and $\epsilon_T$ (bottom) for Trappist-1e DP (left) and FR (right) simulations as function of time during the CME event. Horizontal dashed and dotted lines indicate maximum and initial values. The transfer functions $\epsilon_S$ and $\epsilon_T$ are defined in Eqs. \ref{eq:PoyntingFluxEfficiency} and \ref{eq:TransferFunction}. \label{Fig:TransferFunction_timeseries}}
	\end{figure*} 
	The upper panels of Fig. \ref{Fig:TransferFunction_timeseries} show integrated Poynting fluxes as defined in Eq. \ref{eq:PoyntingFluxInward} for Trappist-1e and planetary magnetic field strengths of $B_p = 0.03$, $0.05$, $0.09$, $0.11$ and $0.21$ G during the CME event. \\

	The CME shock hits the magnetosphere at about 32 minutes and is followed by a sheath region where plasma density, pressure and magnetic flux density is strongly enhanced (see also Fig. \ref{Fig:CME_Structure} for a description of the general CME structure). The sheath crossing ends at approximately 42 minutes with DP CMEs and slightly delayed (1-2 minutes) with FR CMEs. Inward Poynting fluxes rapidly increase by 1 to 2 orders of magnitude with the shock hitting the magnetosphere. For planetary magnetic fields from $B_p = 0.05$ G and stronger $S_{in}^-$ reach powers of $10^{15-16}$ W. $S_{in}^-$ increase with increasing $B_p$, the differences, however, become small for $B_p> 0.05$ G. We note that above $B_p = 0.05$ G the magnetospheres always have magnetopause radii greater that $R_p$ and thus magnetosheath dynamics do not contribute directly to the calculated Poynting fluxes. In DP and FR simulations we observe an oscillatory evolution of $S_{in}^-$ during the first 4 minutes after the shock intercepts the magnetosphere. During this time the CME-induced compression and strong magnetic pressure of the compressed planetary magnetic field counteract each other periodically. 

    In both models the total CME energy flux decays following the exponential decrease of the initial CME profile that is stretched and therefore diluted along the flow direction due to the fast CME front. The magnetopause radius increases due to the decreasing CME ram pressure of the diluted CME tail plasma (Fig. \ref{Fig:magnetopause}). The stronger $B_p$ is, the slower is the decay of $S_{in}^-$ due to the larger area to intercept CME energy.
	\\
	
	For all models strongest Poynting fluxes are generated during the CME sheath and peak crossing with $S_{in}^-$ amounting to $10^{15-16}$ W in a time span of about 10--15 minutes. The convected Poynting flux clearly depends on $B_p$ whereas the dependence becomes stronger when $B_p \ge 0.09$ G. From the beginning of the CME event up to the decay phase (50--90 minutes) $S_{in}^-$ drops by approximately one order of magnitude.
	\\
	\paragraph{\bf{Transfer of energy fluxes}\\}
	We calculate the CME kinetic energy flux, $P_{kin}$, thermal energy flux, $P_{th}$ and Poynting flux, $S_{CME}$, with plasma parameters obtained from our simulations incident on the magnetospheric cross section $\pi R_{mp}^2$,
	\begin{eqnarray}
		P_{kin} &=& \frac{1}{2} \rho v^3 \pi R_{mp}^2 \label{eq:kinEngFlux} \\
		P_{th} &=& \frac{3}{2} n k_B T \pi R_{mp}^2 v \label{eq:thermEngFlux} \\
		S_{CME} &=& \frac{B^2}{\mu_0} \pi R_{mp}^2 v_\perp \label{eq:magEngFlux}\; ,
	\end{eqnarray}
	where $v$ denotes the CME plasma velocity, $v_\perp$ the velocity perpendicular to the magnetic field, $n$ the plasma number density, $T$ the temperature and $B$ the magnetic flux density. All these parameters are obtained directly in front of the magnetosphere where the CME plasma is not yet perturbed. 
	
	We compare the incident CME Poynting flux with $S_{in}^-$ by calculating the Poynting flux ratio 
	\begin{equation}\label{eq:PoyntingFluxEfficiency}
		\epsilon_S = S_{in}^-/S_{CME}\;.
	\end{equation}
	We note that $\epsilon_S$ is not an efficiency factor (that must be $\le 1$) but a fraction that allows us to get an idea of the amount of $S_{in}^-$ that is generated within the magnetosphere due to magnetic field perturbation (if i.e. $\epsilon_S > 1$).
	Additionally, following \citet{Elekes2023}, we calculate the transfer function between total incident energy flux (Eqs. \ref{eq:kinEngFlux}--\ref{eq:magEngFlux}) and $S_{in}^-$, 
	\begin{equation}\label{eq:TransferFunction}
		\epsilon_T = \frac{S_{in}^-}{P_{kin}+P_{th}+S_{CME}}\;.
	\end{equation}
	The transfer function measures the conversion efficiency of incident energy to magnetospheric inward Poynting fluxes. As $\epsilon_T$ relates the total available energy to magnetospheric Poynting fluxes it must be smaller than unity. Together, both quantities may be used to assess the interaction strength and to identify energy transfer that occurs due to mechanical interaction or via reconnection and convection.
	\\
	
	In the middle and bottom panels of Fig. \ref{Fig:TransferFunction_timeseries} we show the time series of $\epsilon_S$ and $\epsilon_T$, respectively. 
	
	In DP model runs, $S_{in}^-$ exceeds the incident Poynting flux by several orders of magnitude. During the CME shock crossing $\epsilon_S$ increases for all $B_p$ by two orders of magnitude up to $10^3$. These high $\epsilon_S$ indicate a high amount of $S_{in}^-$ generated by conversion from CME mechanical to magnetic energy (i.e. by magnetosphere perturbation). The transfer function $\epsilon_T$ reaches a maximum value of $0.4$ during the shock crossing in the $B_p=0.21$ G case and about $0.2$--$0.3$ for $B_p\approx0.05$ G. Afterwards $\epsilon_T$ is reduced to $0.1$ and $0.2$, indicating a slightly lower efficiency of CME energy injection. We thus identify magnetosphere compression to be the dominant mechanism for DP CMEs which is supported by most $dB/dt$ occurring near the polar cusps (Fig. \ref{Fig:MagneticMaps}, left).
	
    \begin{figure*}
		\centering
        \includegraphics[scale=0.246]{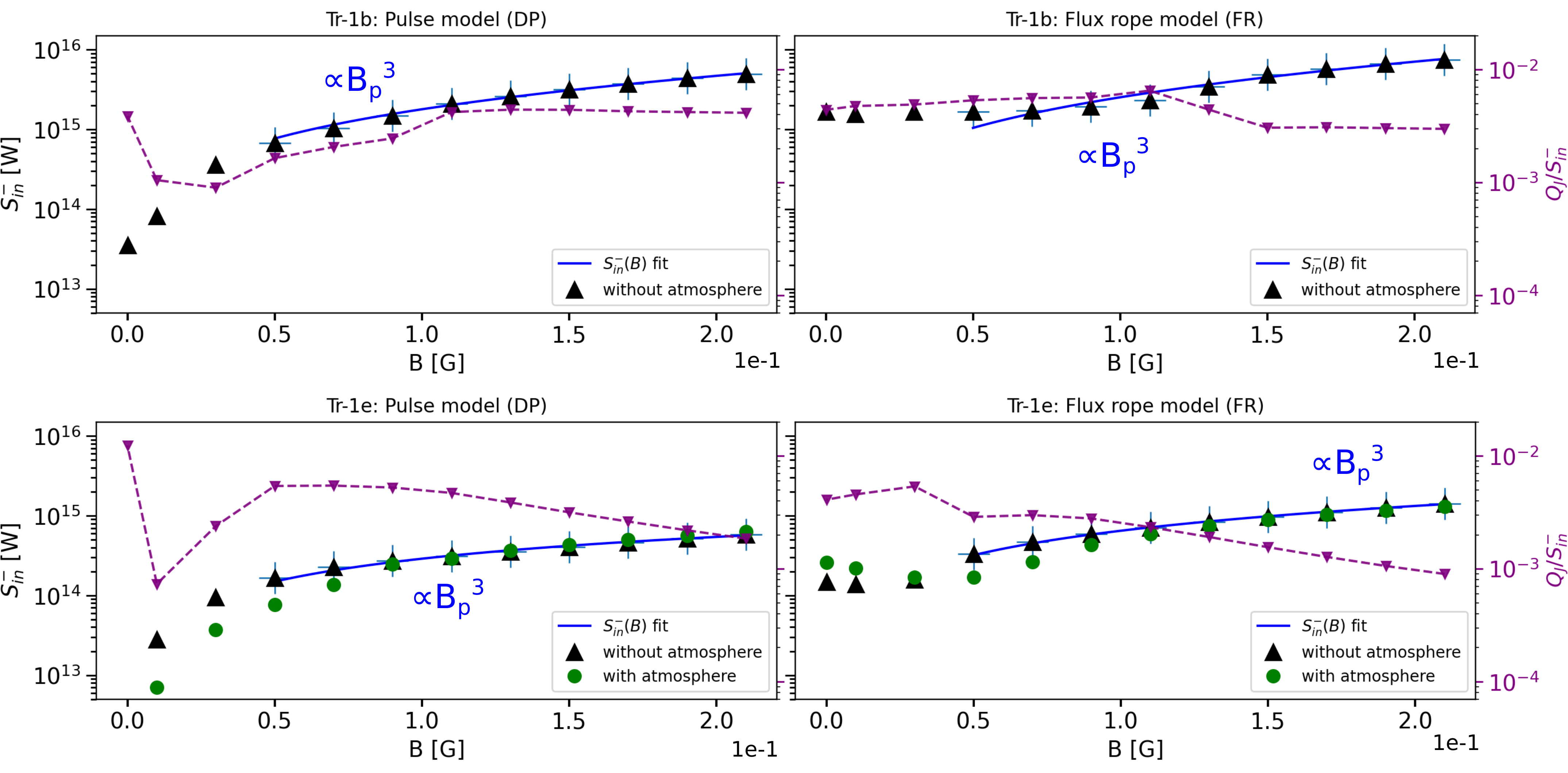}
		\caption{Time-averaged inward Poynting fluxes $S_{in}^-$ in Watts as function of planetary magnetic field flux density $B_p$ in G. DP results are left, FR results are right. The top row shows Trappist-1b and the bottom row Trappist-1e. Black triangles denote $S_{in}^-$. Purple downward triangles denote the ratio of Ohmic heating to inward Poynting flux $Q_J/S_{in}^-$ (right y-axis). The blue solid lines are fits to the Poynting flux data. We excluded the cases where the upstream magnetopause is pushed to the planetary surface during the CME event ($B_p < 0.5$ G). Green circles show $S_{in}^-$ with an atmosphere present. \label{Fig:PoyntingFluxes_average}}
	\end{figure*}
    
	In FR model runs we observe a different behavior of the energy transfer. $\epsilon_S$ is significantly lower, between $1$ and $60$, since magnetosphere compression also occurs in this scenario due to the high velocity of the CME. $\epsilon_T$ is generally higher compared to the DP scenario and lies between $0.1$ and $0.7$. This is due to the strong intrinsic magnetic variability and thus Poynting flux already contained within the FR that dominates the total energy flux. Smaller $\epsilon_S$ during the FR passing indicate that less magnetic energy is released due to a mechanical interaction. This becomes more clear when focusing on the CME sheath and peak passing, where, nearly for all $B_p$, $\epsilon_S$ falls to values between $0.5$ and $3$ while $\epsilon_S$ is smaller for weaker $B_p$ which is supported by the $Q_J$ scaling with weak $B_p$ in Sect. \ref{Sect:InteriorOhmicHeating}. This is indicative for dominant Poynting flux and thus magnetic variability transfer along open field lines due to reconnection. This transfer of magnetic variability is the root cause for the persistent heating efficiency seen in Fig. \ref{Fig:HeatingRates}. If no or a weaker planetary magnetic field is present the magnetic generator mechanism (Eq. \ref{induction-equation}) that converts motional energy into magnetic energy (i.e. $dB/dt$) is not existent or weaker, respectively. For stronger $B_p \ge 0.05$ G magnetic energy released due to compressional perturbation increases, raising $\epsilon_S$ to $2$--$3$.
    
    After the CME peak at about $43$ minutes, $\epsilon_S$ drops off slowly in the DP CME case as the mechanical energy of CME slowly decays in the CME tail. In the FR scenario $\epsilon_S$ shows slight variations due to the flux rope's twisted magnetic field structure slowly breaking down to eventually approach the steady state magnetic field configuration of the stellar wind. For the weaker planetary magnetic field regime ($B_p \le 0.11$ G), $\epsilon_T$ remains fairly constant at $0.1$. In the CME tail region, density and velocity drop to minimum values and in the same time, $R_{mp}$ rises (Fig. \ref{Fig:magnetopause}). The magnetosphere reacts increasingly sensitive to variations in the ambient plasma like density, velocity and magnetic field gradients across the CME tail when the magnetosphere's effective size increases.
	\\
	
	In summary we found evidence for a strongly enhanced CME Poynting flux-to-magnetospheric Poynting flux transfer if the CME possesses a flux rope. Mechanically dominated CMEs convert only a small fraction, $\epsilon_T \le 0.2$ ($0.4$ peak during shock crossing), of the total energy flux to inward Poynting fluxes above the planet surface. Flux rope CMEs already contain large Poynting fluxes, but get them also well into the magnetosphere. With the total energy flux dominated by magnetic energy, the fraction of total energy transferred to magnetospheric Ponyting fluxes increases considerably with $B_p$ and reaches up to $0.7$. To put this value into context, in a steady state stellar wind scenario we found the transfer efficiency to be approximately 0.15--0.2 for the magnetosphere of $\tau$ Boötis b \citep{Elekes2023}.
	\begin{figure*}
		\centering
        \includegraphics[width=1\textwidth]{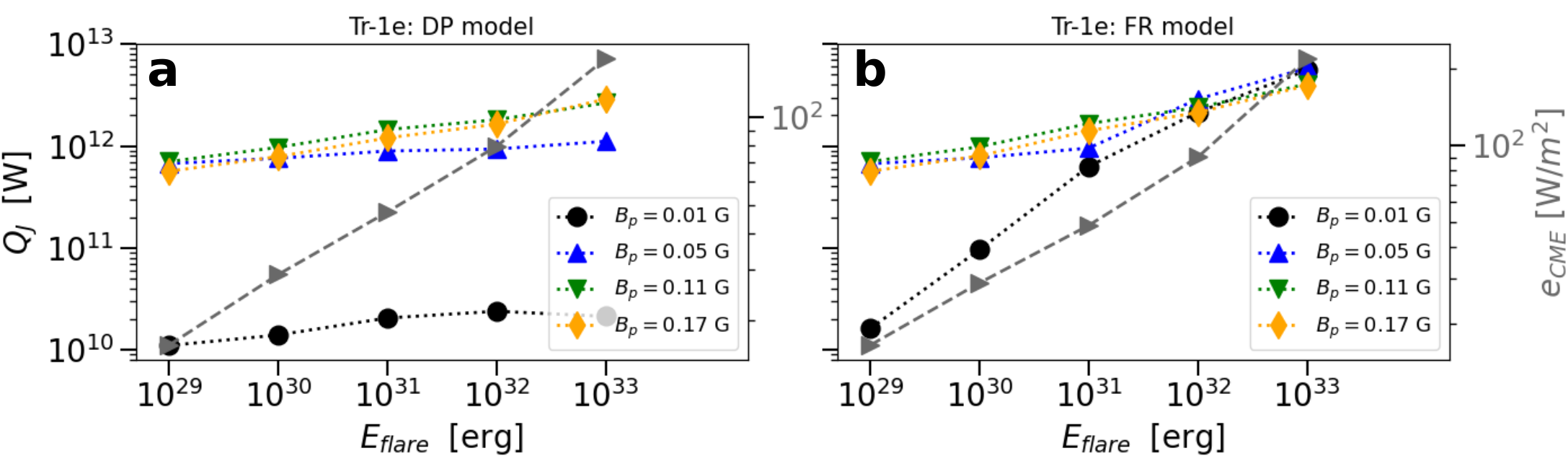}
		\caption{Time-averaged interior heating rates as a function of CME-associated flare energy $E_{flare}$ (erg) for DP (a) and FR (b) CMEs. We only show heating rates for $B_p=0$ G (black circles), $B_p=0.05$ G (blue upward triangles), $B_p=0.11$ G (green downward triangles) and $B_p=0.17$ G (yellow diamonds). Grey triangles denote the incident total energy flux of the CME (right y-axis). \label{Fig:heating_vs_E}}
	\end{figure*}
	\subsubsection{Scaling behavior of magnetospheric Poynting fluxes and the role of planetary magnetic field strength}\label{Sect:Energetics_Scaling}
	We are interested in understanding the functional dependence of $S_{in}^-$ on $B_p$ and thus how planetary magnetic fields affect the Poynting flux above the planet's surface. We average $S_{in}^-$ over the CME duration and display it as a function of $B_p$ for Trappist-1b and e (Fig. \ref{Fig:PoyntingFluxes_average}). Despite the Poynting fluxes varying over orders of magnitudes during the CME event, the average Poynting flux nevertheless quantifies the average energy injected into the magnetosphere during the event. Contrary to the heating rates in Fig. \ref{Fig:HeatingRates} the averaged Poynting fluxes do not have a maximum within our parameter range.
    
    In general, we find that the Poynting flux $S_{in}^-$ always increases with the planetary magnetic field strength $B_p$.
    In the Trappist-1b case $S_{in}^-$ increase from about $10^{13}$ W to $5\times 10^{15}$ W (DP) with the magnetic field $B_p$ while in the FR case the Poynting fluxes exhibit a lower limit of $5\times 10^{15}$ W. For Trappist-1e the Poynting fluxes are significantly reduced by approximately an order of magnitude while the behavior of $S_{in}^-$ as function of $B_p$ is similar to Trappist-1b. 
    As the external forcing (i.e. the CME) is kept constant this scaling of Poynting fluxes with the intrinsic magnetic field illustrates that stronger fields are capable of absorbing and converting more electromagnetic energy, so that the electromagnetic shielding capability of intrinsic magnetic fields is reduced when they become stronger.

    We now fit $S_{in}^-$ with respect to $B_p$ (Fig. \ref{Fig:PoyntingFluxes_average}).
    The results in Fig. \ref{Fig:PoyntingFluxes_average} suggest that inward Poynting fluxes above the planet's surface scale mainly with $B_p$ following a cubic power law. Effectively $S_{in}^-$ scales with $B_p^3$ indicating that stronger intrinsic magnetic fields enhance the magnetosphere's intake of external electromagnetic energy.
	\\
	
	In Fig. \ref{Fig:PoyntingFluxes_average} we also display the ratio of $Q_J$ to $S_{in}^-$ as green, dashed line to assess how the radial Poynting fluxes translate to interior heating as they supply the planetary surface with the majority of magnetic variations. 
    The Poynting flux-to-interior heating efficiency for magnetic fields above $0.05$ G is largest for Trappist-1b, approximately $2$--$3\times 10^{-3}$, and is slightly reduced for Trappist-1e with approximately $10^{-3}$. The efficiency slightly decreases towards an Earth-like magnetic field strength. Peak Poynting flux-to-interior heating efficiencies are found at magnetic field strengths between $0.5$ and $0.11$ G. 
 
	\subsection{Interior heating as a function of CME-associated flare energy}\label{Sect:Heating_vs_E}
    For our basic model we considered a flare bolometric energy of $10^{31}$ erg. Flares of this energy occur on average once per day according to the recent flare frequency distribution of \citet{Howard2023}. Here we address the question of how interior heating relates to flare energy. For this purpose we additionally considered flare energies of $10^{29}$, $10^{30}$, $10^{32}$ and $10^{33}$ erg which all reside in a reasonable energy range for Trappist-1 \citep{Howard2023}. We ran our MHD model with CMEs based on these flare energies (see Sect. \ref{Sect:CME_models} for our CME models and Eqs. \ref{eq:CME_mass}, \ref{eq:CME_velocity} and \ref{eq:Helicity} for flare-CME scaling laws). Unfortunately we were not able to run the same MHD model for more energetic CMEs ($E_{flare}\ge10^{34}$ erg) due to limitations of the numerical method. The CME duration is kept at one hour. In Fig. \ref{Fig:heating_vs_E} we show resulting heating rates as function of flare Energy in erg. For simplicity we only consider Trappist-1e with magnetic fields of $B_p = 0.01$, $0.05$, $0.11$ and $0.17$ G. We also show the CME total energy density as a function of flare energy.

    For CMEs dominated by mechanical energy (DP), $Q_J$ depends relatively weak on flare energy. Heating rates increase by less than an order of magnitude within our flare energy range in the DP case (Fig. \ref{Fig:heating_vs_E}a). In the non-magnetized planet case, $Q_J$ can be considered nearly independent of flare energy since no considerable magnetosphere exists that could act as generator to convert the CME mechanical energy to magnetic energy. With increasing magnetic field strength the magnetospheric generator action enhances and so does the converted magnetic energy that ultimately drives induction in the interior. 
    For flux rope CMEs dominated by magnetic energy (FR) the dependence of interior heating rates on flare energy is fairly equal for all considered planetary magnetic fields as magnetic energy is already contained within the CME and no efficient generator mechanism is needed (Fig. \ref{Fig:heating_vs_E}b). In general, $Q_J(E_{flare})$ follows the increase in incident energy density of FR CMEs with flare energy more closely than in the DP case.
    \\
    
    In comparison to the results of \citet{Grayver2022} our interior heating rates show a weaker functional dependence on flare energy due to the plasma interaction added to the model. When a planet is magnetized, CME energy is never directly transferred towards the planet. A large portion of $dB/dt$ above the planetary surface is produced by induction in the magnetospheric plasma with a certain efficiency, which increases with planetary magnetic field strength (see Sect. \ref{Sect:Energetics_timedomain}). The magnetospheric generator is driven by the CME's mechanical energy and the resulting perturbation of the planetary magnetic field. 
    
    In the weakly magnetized planet case with $B_p=0.01$ G, however, our FR scenario results (Fig. \ref{Fig:heating_vs_E}b) fit well within the distribution of heating rates obtained by \citet{Grayver2022} since the magnetic energy of the flux rope is nearly directly supplied to the planet's interior due to the lack of a significant magnetospheric generator. 
    In the $B_p=0.01$ G case, above $E_{flare}=10^{31}$ erg, $Q_J(E_{flare})$ assumes the functional dependence of the stronger magnetized scenarios (Fig. \ref{Fig:heating_vs_E}b). We account this to the energy partition of the CME kinetic and magnetic energies. The CME velocity grows slower with flare energy compared to the FR magnetic field strength (see Eqs. \ref{eq:CME_mass}--\ref{eq:CME_velocity} and. \ref{eq:Helicity}--\ref{eq:Helicity_GoldHoyle}). From $E_{flare}= 10^{30}$ erg to $10^{33}$ erg the CME velocity enhances by a factor of $1.5$ while the FR magnetic field strength by $18$. Therefore, by decreasing $E_{flare}$ towards $10^{29}$ erg we quickly reach a kinetic energy dominated regime in which the generation of $dB/dt$ mostly depends on the magnetospheric generator action and thus $Q_J(10^{29}\text{erg})$ is nearly the same in the DP and FR scenario.
    
	In summary, in the DP case, $Q_J$ depends less strongly on $E_{flare}$ when the CME is dominated by mechanical energy The increase of $Q_J$ with flare energy enhances with stronger planetary magnetic fields. For CMEs with a magnetic flux rope structure, however, we find a moderately strong dependence that is constant for all intrinsic magnetic fields above approximately $0.03$G.

    \subsection{Long-term interior Joule heating rates}\label{Sect:heating_time_averaged}
    
    \begin{table}
    	\caption{Statistical flare and CME parameters.}      
    	\label{table:average_heating_parameters}      
    	\centering                                   
    	\begin{tabular}{l c c c c c}          
    		\hline\hline                      
    		  $E$$^\text{(erg)}$  &  $10^{29}$ & $10^{30}$ & $10^{31}$  & $10^{32}$ & $10^{33}$ \\ 
    		\hline
            $n_E$ \tablefootmark{$\dagger$} 	& 4$^{+2.1}_{-1.3}$ & 3.6$^{+2.1}_{-1.3}$ & 0.9$^{+1.4}_{-0.5}$ & 0.01$^{+10^{-3}}_{-10^{-3}}$ & 0.001$^{+10^{-4}}_{-10^{-4}}$ \vspace{0.15cm}\\ 
    		$N_E$ \tablefootmark{$\star$} 		& 123$^{+201}_{-85}$ & 110$^{+192}_{-78}$ 	& 28$^{+94}_{-22}$ & 0.3$^{+0.49}_{-0.26}$ & 0.03$^{+0.03}_{-0.02}$ \vspace{0.1cm}\\
    		\hline
   
    	\end{tabular}
    	\tablefoot{
            \tablefoottext{$\dagger$}{Flares with energy $E$ per day $n_E$.\\}
    		\tablefoottext{$\star$}{$N_E = n_E\,f\,\times T$}\\
    		Estimated number of CME events per considered timescale $T$ and flare energy $E_{flare}$ based on the flare frequency distribution of \citet{Howard2023} and \citet{Seli2021}. The averaging timescale is $T=1$ year and the CME event fraction is $f=0.084^{+0.061}_{-0.045}$ \citep{Grayver2022}. See the corresponding Sect. \ref{Sect:heating_time_averaged}.
    		
    	}
    \end{table}

	\begin{table}
		\caption{Time-averaged annual Joule heating rates (TW) in the interior of Trappist-1b \& e due to CMEs.}      
		\label{table:average_heating_rates}      
		\centering                                   
		\begin{tabular}{l c c c c }          
			\hline\hline                       
			$B_p$ (G) &  $0.01$ & $0.05$ & $0.11$  & $0.17$ \\
			\hline

          \textbf{Tr-1b}\tablefootmark{$\dagger$}	 &&&&\\
            DP& 	         & 0.61$^{+1.2}_{-0.4}$ &  3.3$^{+6.5}_{-2.4}$ 	& 8$^{+15.8}_{-5.7}$ 	\vspace{0.15cm}\\
			FR&  	         & 3.5$^{+7.1}_{-2.5}$	    &  6.4$^{+12.8}_{-4.6}$ 	    & 7.1$^{+14.2}_{-5.1}$      \vspace{0.1cm}\\		
			\hline\hline
			 \textbf{Tr-1e}	&&&&\\
            DP& 0.009$^{+0.02}_{-0.01}$    & 0.52$^{+1}_{-0.4}$  &  0.64$^{+1.3}_{-0.5}$  & 0.52$^{+1}_{-0.4}$  \vspace{0.15cm} \\
			FR& 0.08$^{+0.2}_{-0.1}$ 	 & 0.53$^{+1}_{-0.4}$ 	&  0.66$^{+1.3}_{-0.5}$ 	& 0.54$^{+1.1}_{-0.4}$	\vspace{0.1cm}\\		

			\hline

		\end{tabular}
		\tablefoot{
			Annual average heating rates in TW are calculated using Eq. \ref{eq:heating_rate_year}, the parameters in Table \ref{table:average_heating_parameters} and the single-event heating rates for different CME-associated flare energies (Sect. \ref{Sect:Heating_vs_E}). \\
            \tablefoottext{$\dagger$}{Underlying $Q_J(E)$ was estimated from Tr-1e case and not obtained from simulations.}
		}
	\end{table}

    Using the flare frequency distributions of Trappist-1 and analogue stars \citep{Howard2023,Seli2021} together with single-event Joule heating rates for different CME-associated flare energies (Sect. \ref{Sect:Heating_vs_E}, Fig. \ref{Fig:heating_vs_E}) we estimate the time-averaged heating rate over one year with several different CME events. Given the average number of flare events per day $n_E(E_i)$ for a given flare energy $E_i$ taken from \citet{Howard2023} we can estimate the average number $N_E$ of flares per year for each flare energy, $N_E=n_E \cdot 365$. We divide the flare energy range into logarithmically spaced energy bins, $E_i \in [10^{29},10^{30},10^{31},10^{32},10^{33}]$ erg, with $i \in [0,4]$. Single event heating rates derived with our CME model are shown in Fig. \ref{Fig:heating_vs_E}. The fraction of all flare events that are associated with a CME which intersects a planet is $f\approx 0.08$ according to the estimate of \citet{Grayver2022}. 
    We multiply the heating rates $Q_E(E_i)$ due to CMEs with energy $E_i$ (Fig. \ref{Fig:heating_vs_E}) with their corresponding number of annual events and then divide them by $N=365$ days to obtain an estimate for the average annual heating rate, $Q_{CME}$, 
    \begin{equation}\label{eq:heating_rate_year}
        Q_{CME} = \frac{f}{N}   \sum_{i}^{}  N_E(E_i) \cdot Q_E(E_i)  \;.
    \end{equation}
    Resulting annual average heating rates are displayed in the lower box in Table \ref{table:average_heating_rates}. We also estimated the annual heating rates for Tr-1b by assuming $Q_J(E)$ having the same form as for Tr-1e (\ref{Fig:heating_vs_E}). We scaled the fit function in Fig. \ref{Fig:heating_vs_E} to match the heating rate $Q_J(E=10^{31}$ erg$)$ obtained from our basic model for Tr-1b (Sect. \ref{Sect:InteriorOhmicHeating}, Fig. \ref{Fig:HeatingRates}). With $Q_J(E)$ obtained we applied the same calculations as for Tr-1e presented in this section. We omitted the data for $B_p=0.01$ G due to the substantially different functional dependence of $Q_J$ on $E$ in this case (Fig. \ref{Fig:heating_vs_E}).

    Large scale stellar magnetic fields are capable of suppressing or slowing down CMEs, leading to lower energies contained in CMEs intersecting planets (see Sect. \ref{Sect:Uncertainties}). We therefore note that the dissipation rates presented here are presumably upper limits.
    Annual interior heating due to CMEs are on the order of $1$--$10$ TW for Trappist-1b and $0.1$--$1$ TW for Trappist-1e. Comparing these results to the heating rates of \citet{Grayver2022} obtained from a homogeneous interior model we find our heating rates to lie within the lower range of their results. We find the best agreement between our results and those of \citet{Grayver2022} for very weakly magnetized planets. Therefore, by considering a CME model incorporating time-dependent MHD simulations of the CME-planet interaction, CME induced interior Joule heating has a tendency towards lower dissipation rates. Our heating rates for Trappist-1e are insignificant compared to the tidal heating rates estimated by \citet{Bolmont2020} which are, on average, 1--2 orders of magnitude higher. Compared to the interior heating rates due to the planet's motion through the stellar dipole field \citep[][corrected by \citet{Grayver2022} using a more recent stellar rotation rate]{Kislyakova2017} our results are roughly one order of magnitude lower. However, the results of \citet{Kislyakova2017} depend on the unconstrained tilt of its magnetic moment with respect to the planet's orbital plane and thus, with low inclination the resulting heating rates are less.

	\section{Summary}
	In this study, we investigated the interaction between CMEs and exoplanets using MHD simulations. We chose Trappist-1b and e as an exemplary case, and studied the resulting magnetic variability on the surface of the planets. CME-generated magnetic variations at the surface of the planets induce electric currents in the planetary interior, which dissipate energy in the form of heat. We have developed an MHD model to simulate the interaction of density pulse (DP) and flux rope (FR) CMEs with magnetized and non-magnetized planets. With the choice of magnetically (FR) and mechanically (DP) dominated CMEs we are able to study in detail the transfer of CME energy towards the planet's surface separately. This way we are able to better understand the role of planetary magnetic fields in the conversion of energy and in governing dissipation of external energy fluxes within the planet and its atmosphere.
    We considered a range of intrinsic dipolar magnetic fields from $B_p=0$ G to a nearly Earth-like field with $B_p=0.21$ G. We derived CME parameters from flare-CME scaling laws obtained from Solar system observations and applied them to CMEs with associated flare energies from $E_{flare} = 10^{29}$ to $10^{33}$ erg. Our CMEs have a duration of $1$ hour. We used a planetary interior model with constant electric conductivity and calculated induction heating using the model of \citet{Grayver2022}. For Trappist-1e we also considered an O$_2$ atmosphere and calculated ionospheric Joule heating during the CME event. In the following we summarize the main results of this study.\\

	\textbf{1)} Taking the MHD processes around the planets into account, we find that CME Induction heating in the interior is smaller and within the lower end of the heating rates previously derived \citet{Grayver2022}. 
    Within the parameter space studied here, an intrinsic magnetic field generally acts as an "enhancer" of stellar wind magnetic variability on the surface of the planets which confirms the results of \citet{Grayver2022}.
    Stresses on the magnetosphere are argued to be the main driver of magnetic variability at the planetary surface. For flux rope CMEs and with weak planetary $B_p$ the intrinsic magnetic variability of the CME can be directly transported to the planet's surface. With increasing $B_p$, however, the magnetic energy that can be released by mechanical perturbation exceeds the magnetic energy of the flux rope.
    Interior heating due to flux rope CMEs is enhanced when an atmosphere is present due to additional tension imposed on the reconnected magnetic field lines.

    \textbf{2)} Ionospheric Joule heating rates range from $1$--$4\times10^{4}$ TW for weakly magnetized planets down to $1$--$3\times10^{3}$ TW towards stronger magnetic fields. Joule heating of the upper atmosphere does not rely on electromagnetic induction but directly results from the increase of magnetospheric electric fields. With this nearly instantaneous heating during CMEs atmospheric inflation and possible erosion might be drastically enhanced. CME induced Joule heating rates in the upper atmosphere exceed the XUV radiation by 1--2 orders of magnitude. These large dissipation rates suggest that a single CME might have severe effects on the atmosphere of Trappist-1e and possibly similar close-in terrestrial exoplanets. Ionospheric Joule heating rates during the steady state stellar wind amount to $\approx 2\times10^2$ TW if the planet is strongly magnetized and up to $8\times10^3$ TW without the presence of a planetary magnetic field.
	
	\textbf{3)} Regions on the planetary surface with strong radial $dB/dt$ are increasingly confined to areas near the upstream polar cusps within the closed field line region for stronger planetary magnetic fields. Inward oriented radial Poynting fluxes $S_{in}^-$ at the planetary surface coincide spatially well with the $dB/dt$ maxima.

    \textbf{4)} Our results suggests that, in an electromagnetic sense, planetary magnetic fields enhance the capability of planets to receive and convert the energy injected through CMEs. The generation and transfer of field-aligned Poynting fluxes towards the planet as well as inductive coupling between magnetospheric variations and the planet's interior correlates well with increasing intrinsic magnetic field strength. The dependence of this behavior is only moderately sensitive to changes in CME energy density if the planet is at least weakly magnetized. Planetary magnetic fields do not shield the planet's surface from electromagnetic energy received from CMEs and other bursty stellar wind variations as long as the planet does not posses a highly conductive ionosphere.

	\textbf{5)} We find a generally weak functional dependence of interior heating rates $Q_J$ on CME-associated flare energy within our parameter space. For density pulse (DP) CMEs the dependence is almost non-existent for weak planetary magnetic fields $B_p \lesssim 0.01$ G). For stronger magnetic fields, heating rates increase by less than an order of magnitude within our considered flare energy range. For $B_p \gtrsim 0.1$ G the scaling seems to saturate.
    For FR CMEs heating rates increase by almost one order of magnitude within the flare energy range considered. The dependence of heating rates on flare energy is approximately constant for all considered planetary magnetic fields $B_p > 0.01$ G.
    Heating rates scale strongly with $E_{flare}$ for very weakly magnetized planets with $B_p\le0.01$ G.

    \textbf{6)} From the flare frequency distribution of the Trappist-1 star we estimated the average number of annual CME events. Together with our simulated interior heating rates due to CMEs with different flare energies we calculated the annual average interior heating rates to be on the order of $1-10$ TW (Tr-1b) and $0.1-1$ TW (Tr-1e). These heating rates are approximately 2 orders of magnitude lower compared to the results of \citet{Grayver2022}. Thus, considering an MHD model of the CME-planet interaction we find CME-induced interior Joule heating to be less significant compared to dissipation rates obtained from purely electromagnetic models due to the highly nonlinear coupling between external forcing and interior dissipation. Our estimated average dissipation rates are about 2 orders of magnitude lower than estimates for tidal heating in Trappist-1b \citep{Bolmont2020}.
	
	\begin{acknowledgements}
		This project has received funding from the European Research Council (ERC) under the European Union’s Horizon 2020 research and innovation programme (grant agreement No. 884711).
		\\
		The authors gratefully acknowledge the Gauss Centre for Supercomputing e.V. (www.gauss-centre.eu) for funding this project by providing computing time through the John von Neumann Institute for Computing (NIC) on the GCS Supercomputer JUWELS at Jülich Supercomputing Centre (JSC).
		\\
		This work used resources of the Deutsches Klimarechenzentrum (DKRZ) granted by its Scientific Steering Committee (WLA) under project ID 1350.
		\\
		We furthermore thank the Regional Computing Center of the University of Cologne (RRZK) for providing computing time on the DFG-funded (Funding number: INST 216/512/1FUGG) High Performance Computing (HPC) system CHEOPS as well as support.
        \\
        This work was partially funded from the Dutch Research Council (NWO), with project number VI.C.232.041 of the Talent Programme Vici.
	\end{acknowledgements}
	
	\bibliographystyle{aa}
	\bibliography{bib/Tr1_article}
	\clearpage

    \begin{appendix}  
    \section{Photo-ionization rate and atmospheric mass loss}\label{Sect:AtmMassLoss}
    In order to estimate the photo-ionization rate $\beta_{ph}$ which results in a given atmospheric mass loss rate $\dot{M}$ of a certain particle species, we consider the following equation that relates mass loss to the ionization rate,
    \begin{equation}\label{Eq:Ch5:MassLossRate}
	   \dot{M}_{\text{O}_2} = m_{\text{O}_2}\, \beta_{ion}  \int_{V}^{} \int_{z}^{}  n_{\text{O}_2}(r) \,  d\text{V}\; ,
    \end{equation}
    where $m_{\text{O}_2}$ is the mass of the species (here molecular oxygen), $n_{\text{O}_2}(r)$ the atmospheric $\text{O}_2$ number density as a function of radial distance $r$ from the planet and $V$ the volume of the atmosphere. The O$_2$ number density of the neutral atmosphere, $n_{\text{O}_2}(r)$, is given by our model (Eq. \ref{eq:atmosphere}). Only a thin atmosphere is considered in our MHD model with a maximum neutral number density of $8\times10^{12}$ $m^{-3}$ and thus, for the sake of simplicity we neglect the optical depth so that all particles within the atmosphere are photo-ionized at the same rate.
    We integrate Eq. \ref{Eq:Ch5:MassLossRate} analytically from the planetary surface at $R_p$ to a height $z$ (i.e. top of the atmosphere), so that
    \begin{align}\label{Eq:Ch5:MassLossRate_integral}
	   \dot{M}_{\text{O}_2} &= m_{\text{O}_2}\,n_{O_2,0}\, \beta_{ion} \int_{R_p}^{R_p+z} \int_{0}^{\pi} \int_{0}^{2\pi} r^2 \sin{\theta}  \exp{\left(\frac{R_p - r}{H}\right)} d\text{r}d\text{$\theta$}d\text{$\phi$} \\
        \label{Eq:Ch5:MassLossRate_solved}
         &\approx m_{\text{O}_2}\,n_{O_2,0}\, \beta_{ion}\,4\pi\, H\, R_p^2\;,
    \end{align}
    where $H$ is the atmosphere scale height.
    Equation \ref{Eq:Ch5:MassLossRate_integral} can be rearranged to yield the desired photo-ionization rate $\beta_{ion}$.
    \section{CME profiles}\label{Sect:Profiles}
     
\begin{figure*}
		\centering
        \includegraphics[width=0.99\linewidth]{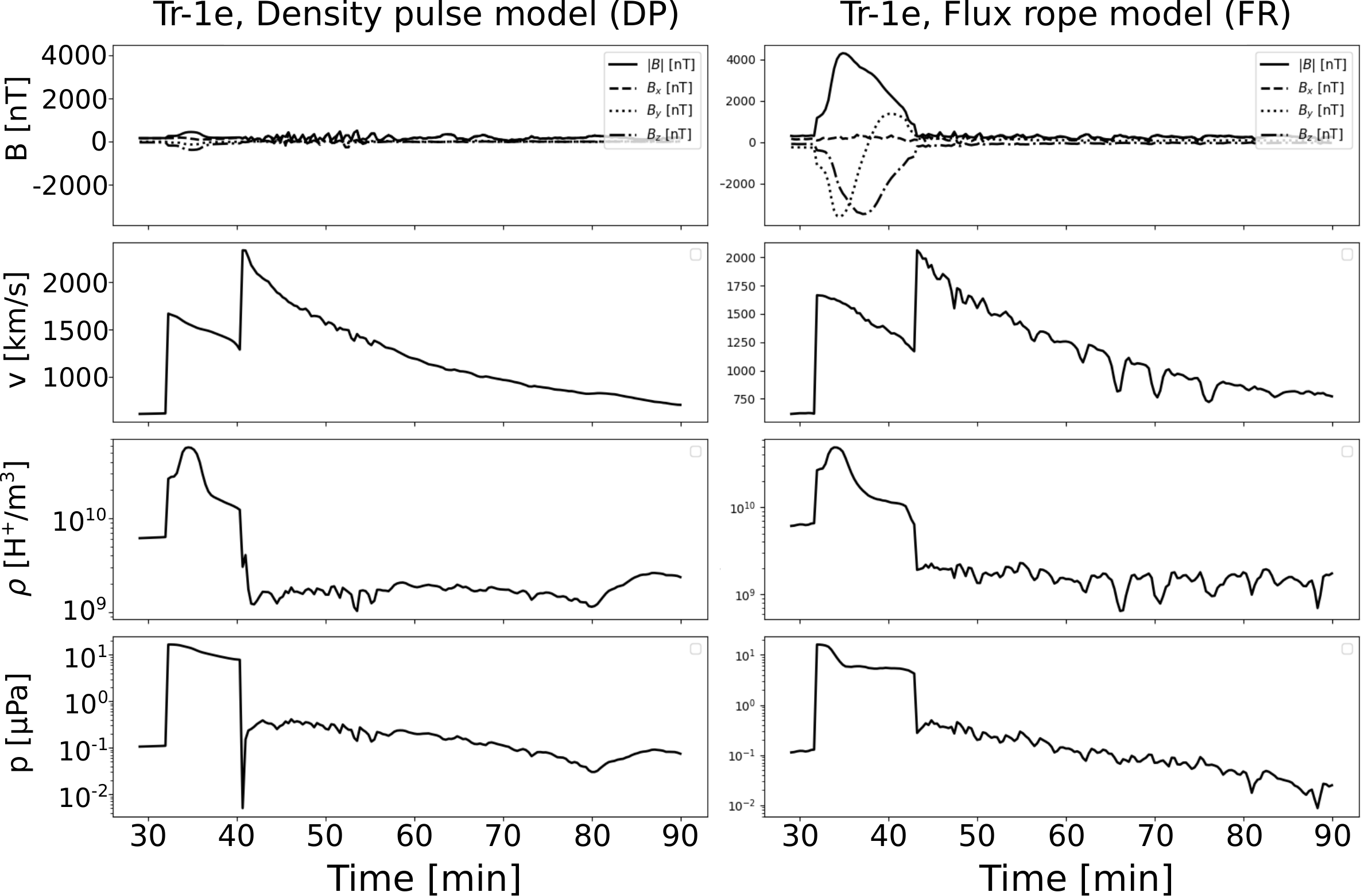}
		\caption{Profiles of all plasma quantities during the CME event (DP left, FR right) of our basic model ($10^{31}$ erg) for Trappist-1e. The profiles are obtained in front of the planetary bow shock letting time evolve. \label{Fig:profiles}}
	\end{figure*}

    In order to illustrate the temporal variation of the MHD variables, i.e. magnetic field, velocity, ion mass density and thermal pressure, we show corresponding profiles in Fig. \ref{Fig:profiles} for Trappist-1e during the CME event. The profiles are obtained from a fixed point in front of the planetary bow shock letting time evolve.
    %
    %
    %
    \section{Influence of CME event duration on interior Joule heating}\label{Sect:DurationInfluence}
    \begin{figure}
		\centering
		\includegraphics[width=0.99\linewidth]{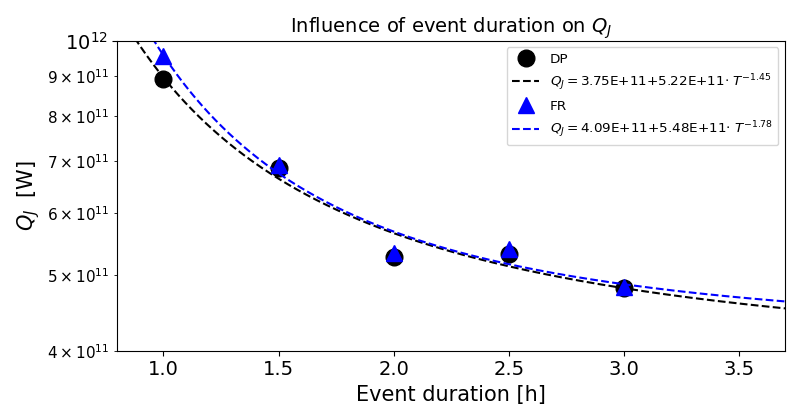} 
		\caption{Interior heating rates as a function of CME event duration (hours). Circles depict DP and triangles FR results. Dashed lines show the fitted curves with the fitting parameters shown in the legend. \label{Fig:heating_vs_T}}
	\end{figure}
    
    In the solar system CME events typically last for several hours up to few days. In order to assess the effect of event duration on the results of this work, we additionally performed simulations with CME event durations up to 3 hours.
    
    We restricted the additional simulations to the DP CME scenario ($10^{31}$ erg) with $B_p=0.05$ G and event duration of $1$ h, $1.5$ h, $2$ h, $2.5$ h and $3$ h. We extended the CME peak (DP model) in form of a plateau with maximum amplitude from the CME center towards the front in such a way so that the front of the curve (i.e. the nose of the Gaussian) has the same shape as in our basic model. With this setup the shape of the shock and sheath region of the CME remain the same. The tail of the CME is, however, stretched out accordingly to match the desired event duration. We find in our results that the heating rates for a single event decrease towards longer event durations (Fig. \ref{Sect:DurationInfluence}). This is caused by the slower decay of the CME parameters in the tail and by the longer averaging period for $dB/dt$. The magnetosphere weakly pulsates during the sustain phase in which we extend the CME peak. This is caused by compression and repulsive expansion of the dayside magnetosphere and thus the extended peak still leads to magnetic variability near the planet although of small amplitude. We fit the heating rates as a function of event duration (dashed lines in Fig. \ref{Sect:DurationInfluence}). Extrapolating $Q_J$ towards longer CME durations we find a saturation of interior heating at about half the value of our basic one hour CME model. We note that this result is likely a consequence of our CME model choices and we do not derive a general trend for CMEs from it but point out that, in our CME model, the event duration only weakly influences the results presented in this work.

	\section{Poynting flux maps}\label{Sect:S_map}
    \begin{figure*}[t]
		\centering
		\includegraphics[width=0.99\linewidth]{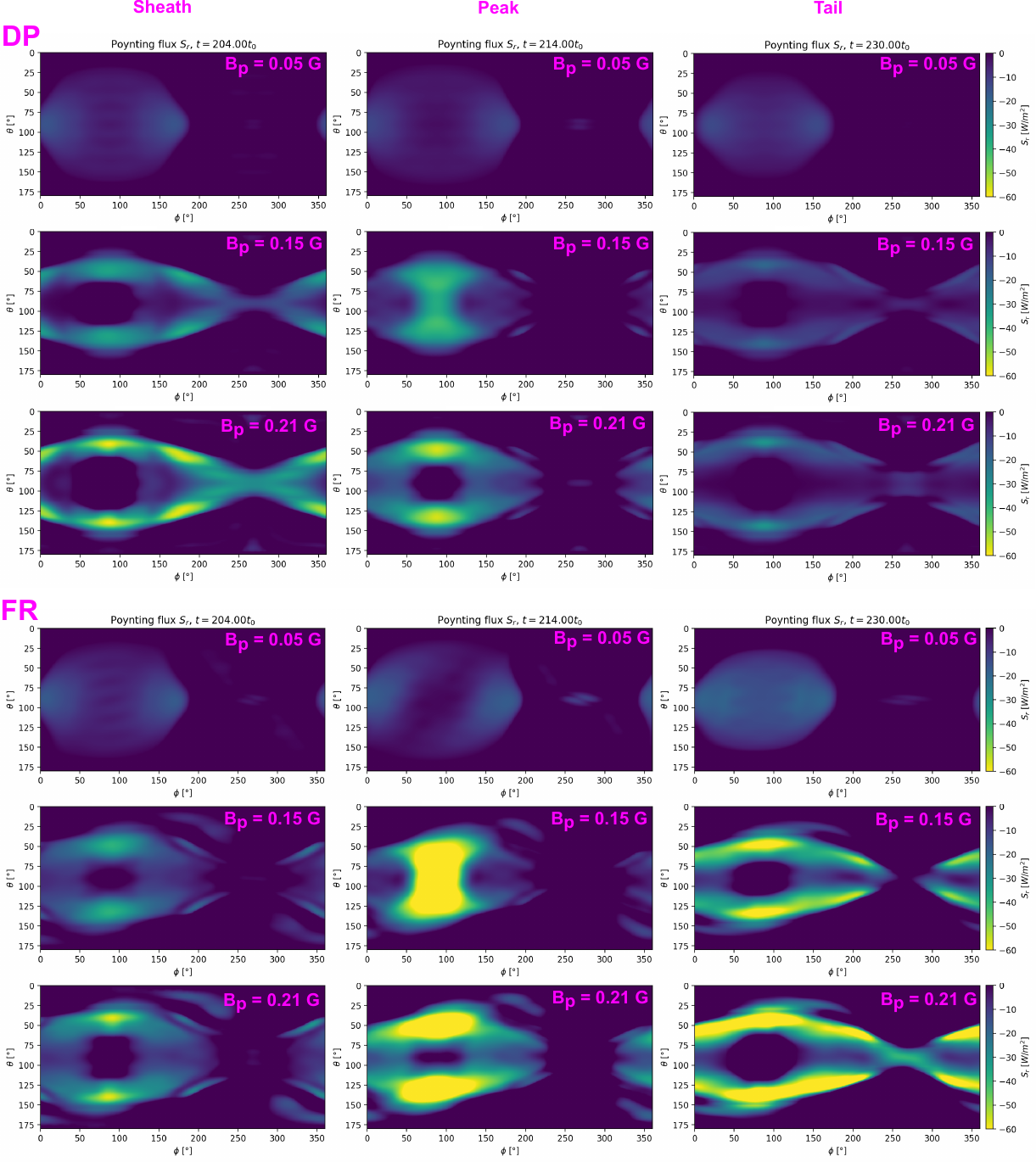} 
		\caption{Maps of inward, radial Poynting flux components, $S_r$, above the surface of Trappist-1e in the DP (top) and FR (bottom) scenario. Maps are shown during the CME sheath (left), peak (middle) and tail crossing (right) (see also Fig. \ref{Fig:CME_Structure}). \label{Fig:S_map}}
	\end{figure*}
    
    Here we present maps of inward Poynting flux components $S_{in}^-$ obtained at the planet's surface (Fig. \ref{Fig:S_map}) complementary to the discussion in Sect. \ref{Sect:Energetics}, Fig. \ref{Fig:ocfb}. The maps demonstrate the well spatial alignment of the Poynting fluxes with the time-averaged magnetic variability of the radial field component $dB/dt$ (Fig. \ref{Fig:MagneticMaps}). The distribution of radial Poynting fluxes support our conclusion that magnetic variability is dominantly generated by magnetospheric compression due to the increased occurrence of $S_{in}^-$  within the closed magnetosphere (i.e. at low latitudes).

    \end{appendix}
\end{document}